\newif\ifcomments\commentsfalse
\newif\ifanon\anonfalse
\newif\iffinal\finalfalse
\newif\ifextended\extendedtrue

\documentclass[10pt,sigplan,screen,dvipsnames\ifanon,anonymous\fi]{acmart}

\iffinal
\else
\fi

\usepackage{microtype}
\usepackage{prettyref}
\usepackage{float}
\usepackage{ragged2e}   

\newcommand{\pref}{\prettyref}
\newcommand{\Pref}{\prettyref}  

\newrefformat{fig}{Figure~\ref{#1}}  

\floatstyle{boxed}
\newfloat{floatbox}{t}{box}
\floatname{floatbox}{Box}
\newrefformat{box}{Box~\ref{#1}}

\ifcomments
\newcommand{\rae}[1]{\textcolor{magenta}{RAE: #1}}
\newcommand{\ab}[1]{\textcolor{blue}{AB: #1}}
\else
\newcommand{\rae}[1]{}
\newcommand{\ab}[1]{}
\fi

\iffinal
\setcopyright{acmcopyright}
\copyrightyear{2018}
\acmYear{2018}
\acmDOI{XXXXXXX.XXXXXXX}

\acmPrice{15.00}
\acmISBN{978-1-4503-XXXX-X/18/06}

\acmConference[Haskell'22]{Haskell Symposium}{15--16 September,
  2022}{Ljubljana, Slovenia}

\else 
\setcopyright{none}
\fi

\makeatletter
\@ifundefined{lhs2tex.lhs2tex.sty.read}%
  {\@namedef{lhs2tex.lhs2tex.sty.read}{}%
   \newcommand\SkipToFmtEnd{}%
   \newcommand\EndFmtInput{}%
   \long\def\SkipToFmtEnd#1\EndFmtInput{}%
  }\SkipToFmtEnd

\newcommand\ReadOnlyOnce[1]{\@ifundefined{#1}{\@namedef{#1}{}}\SkipToFmtEnd}
\usepackage{amstext}
\usepackage{amssymb}
\usepackage{stmaryrd}
\DeclareFontFamily{OT1}{cmtex}{}
\DeclareFontShape{OT1}{cmtex}{m}{n}
  {<5><6><7><8>cmtex8
   <9>cmtex9
   <10><10.95><12><14.4><17.28><20.74><24.88>cmtex10}{}
\DeclareFontShape{OT1}{cmtex}{m}{it}
  {<-> ssub * cmtt/m/it}{}

\DeclareFontShape{OT1}{cmtt}{bx}{n}
  {<5><6><7><8>cmtt8
   <9>cmbtt9
   <10><10.95><12><14.4><17.28><20.74><24.88>cmbtt10}{}
\DeclareFontShape{OT1}{cmtex}{bx}{n}
  {<-> ssub * cmtt/bx/n}{}

\newcommand{\anonymous}{\kern0.06em \vbox{\hrule\@width.5em}}

\newcommand{\bind}{\mathbin{>\!\!\!>\mkern-6.7mu=}}

\usepackage{polytable}

\@ifundefined{mathindent}%
  {\newdimen\mathindent\mathindent\leftmargini}%
  {}%

\def\resethooks{%
  \global\let\SaveRestoreHook\empty
  \global\let\ColumnHook\empty}
\newcommand*{\savecolumns}[1][default]%
  {\g@addto@macro\SaveRestoreHook{\savecolumns[#1]}}
\newcommand*{\restorecolumns}[1][default]%
  {\g@addto@macro\SaveRestoreHook{\restorecolumns[#1]}}
\newcommand*{\aligncolumn}[2]%
  {\g@addto@macro\ColumnHook{\column{#1}{#2}}}

\resethooks

\newcommand{\onelinecommentchars}{\quad-{}- }
\newcommand{\commentbeginchars}{\enskip\{-}
\newcommand{\commentendchars}{-\}\enskip}

\newcommand{\visiblecomments}{%
  \let\onelinecomment=\onelinecommentchars
  \let\commentbegin=\commentbeginchars
  \let\commentend=\commentendchars}

\newcommand{\invisiblecomments}{%
  \let\onelinecomment=\empty
  \let\commentbegin=\empty
  \let\commentend=\empty}

\visiblecomments

\newlength{\blanklineskip}
\newcommand{\hsindent}[1]{\quad}
\let\hspre\empty
\let\hspost\empty

\EndFmtInput
\makeatother
\ReadOnlyOnce{polycode.fmt}%
\makeatletter

\newcommand{\hsnewpar}[1]%
  {{\parskip=0pt\parindent=0pt\par\vskip #1\noindent}}

\newcommand{\hscodestyle}{}

\newcommand{\sethscode}[1]%
  {\expandafter\let\expandafter\hscode\csname #1\endcsname
   \expandafter\let\expandafter\endhscode\csname end#1\endcsname}

  {\par\noindent
   \advance\leftskip\mathindent
   \hscodestyle
   \let\\=\@normalcr
   \let\hspre\(\let\hspost\)%
   \pboxed}%
  {\endpboxed\)%
   \par\noindent
   \ignorespacesafterend}

  {\hsnewpar\abovedisplayskip
   \advance\leftskip\mathindent
   \hscodestyle
   \let\hspre\(\let\hspost\)%
   \pboxed}%
  {\endpboxed%
   \hsnewpar\belowdisplayskip
   \ignorespacesafterend}

  {\hsnewpar\abovedisplayskip
   \advance\leftskip\mathindent
   \hscodestyle
   \let\\=\@normalcr
   \(\pboxed}%
  {\endpboxed\)%
   \hsnewpar\belowdisplayskip
   \ignorespacesafterend}

\newcommand{\plainhs}{\sethscode{plainhscode}}

\plainhs

  {\hsnewpar\abovedisplayskip
   \advance\leftskip\mathindent
   \hscodestyle
   \let\\=\@normalcr
   \(\parray}%
  {\endparray\)%
   \hsnewpar\belowdisplayskip
   \ignorespacesafterend}

  {\parray}{\endparray}

  {\(\parray}{\endparray\)}

\def\codeframewidth{\arrayrulewidth}
\RequirePackage{calc}

  {\parskip=\abovedisplayskip\par\noindent
   \hscodestyle
   \arrayrulewidth=\codeframewidth
   \tabular{@{}|p{\linewidth-2\arraycolsep-2\arrayrulewidth-2pt}|@{}}%
   \hline\framedhslinecorrect\\{-1.5ex}%
   \let\endoflinesave=\\
   \let\\=\@normalcr
   \(\pboxed}%
  {\endpboxed\)%
   \framedhslinecorrect\endoflinesave{.5ex}\hline
   \endtabular
   \parskip=\belowdisplayskip\par\noindent
   \ignorespacesafterend}

\newcommand{\framedhslinecorrect}[2]%
  {#1[#2]}

  {\(\def\column##1##2{}%
   \let\>\undefined\let\<\undefined\let\\\undefined
   \newcommand\>[1][]{}\newcommand\<[1][]{}\newcommand\\[1][]{}%
   \def\fromto##1##2##3{##3}%
   }{\) }%

  {\let\orighscode=\hscode
   \let\origendhscode=\endhscode
   \def\endhscode{\def\hscode{\endgroup\def\@currenvir{hscode}\\}\begingroup}
   \orighscode\def\hscode{\endgroup\def\@currenvir{hscode}}}%
  {\origendhscode
   \global\let\hscode=\orighscode
   \global\let\endhscode=\origendhscode}%

\makeatother
\EndFmtInput
\definecolor{BlueViolet}{HTML}{473992}
\definecolor{OliveGreen}{HTML}{3C8031}
\definecolor{Sepia}{HTML}{671800}

\newcommand{\keyword}[1]{\textcolor{BlueViolet}{\textbf{#1}}}
\newcommand{\id}[1]{\textsf{\textsl{#1}}}
\newcommand{\varid}[1]{\textcolor{Sepia}{\id{#1}}}
\newcommand{\conid}[1]{\textcolor{OliveGreen}{\id{#1}}}

\newcommand{\package}[1]{\href{http://hackage.haskell.org/package/#1}{\textsf{#1}}}

\newcommand{\ext}[1]{\texttt{#1}}

\ReadOnlyOnce{forall.fmt}%
\makeatletter

\let\HaskellResetHook\empty
\newcommand*{\AtHaskellReset}[1]{%
  \g@addto@macro\HaskellResetHook{#1}}
\newcommand*{\HaskellReset}{\HaskellResetHook}

\newcommand\hsforall{\global\let\hsdot=\hsperiodonce}
\newcommand\hsexists{\global\let\hsdot=\hsperiodonce}
\newcommand*\hsperiodonce[2]{#2\global\let\hsdot=\hscompose}
\newcommand*\hscompose[2]{#1}

\AtHaskellReset{\global\let\hsdot=\hscompose}

\HaskellReset

\makeatother
\EndFmtInput

\def\eiger/{\textsc{Eiger}}
\def\Eiger/{\eiger/}

\begin{document}

\title{\Eiger/: Auditable, executable, flexible legal regulations}

\author{Alexander Bernauer}
\email{alexander.bernauer@pwc.ch}
\affiliation{%
  \institution{PwC Switzerland}
  \streetaddress{Birchstrasse 160}
  \city{Zurich}
  \country{Switzerland}
  \postcode{8050}
}

\author{Richard A.~Eisenberg}
\email{rae@richarde.dev}
\affiliation{%
  \institution{Tweag}
  \streetaddress{207 Rue de Bercy}
  \city{Paris}
  \country{France}
  \postcode{75012}
}

\begin{abstract}
Despite recent advances in communication and automation, regulations are still written in natural-language prose, subject to ambiguity, inconsistency, and incompleteness.
How can we craft regulations with precision?
Our solution is embodied in \eiger/, a domain-specific programming language embedded in Haskell.
A domain expert pairs with a software engineer to write regulations in \eiger/.
The domain expert needs only to \emph{read} and \emph{audit} the code, but not write it.
A first, limited, user study suggests that this works well in practice because \eiger/ code mostly looks like Excel formulas with simple SQL queries.

\Eiger/ forms the kernel of a new strategy to deliver value to clients in our professional services business with increased automation and precision.
The framework is \emph{executable}:
with client data, we can use \eiger/ both to deduce how best to adapt to a new regulation and then maintain compliance.
This paper reviews the design of \eiger/ and walks through its implementation.
To preserve a straightforward surface syntax but with monadic semantics, we have leveraged advanced features, including \ensuremath{\conid{\conid{GHC}.Generics}}, the new \ext{Over\-load\-ed\-Rec\-ord\-Dot} extension, and a novel approach to performing class instance selection at run-time.
\end{abstract}

\maketitle

\section{Regulations are \emph{so} 20th century}
\label{sec:intro}
In 2011, Marc Andreessen famously realized that ``software is eating the world''~\cite{software-is-eating-the-world}, transforming business, government, and private life.
Yet the domain of legislation and regulation has remained largely unaffected by this revolution.
Thought leaders have been explaining why the tax domain, for example, is ``late'' and why it is ``next'' \cite{ulbrich-taxman-superpowers}.
Triggered by the global financial crisis of 2007--2008 and fueled by ``rising pressure to secure revenues'' and the ``challenge oactual f supervising and serving an increasingly complex and fast-paced world'', authorities have responded with more and more complex laws and regulations and with ``significant [investments] to develop new capabilities''.

However, society has not yet reaped all expected returns: regulation drafting remains a manual, ambiguous, error-prone process, and regulators still have insufficient insight into systemic risks.
Instead, regulated entities must each ensure compliance separately, duplicating the work necessary to convert regulation text into an executable specification, incurring high costs in the process~\cite{cost-of-compliance}.
Other consequences include high mitigation costs for authorities, frequent lawsuits to address the deficits of the legal prose, as well as a general corrosion of the rule of law.

So, imagine instead a future when legislation in available in prose \emph{and} in code.
Critically, this code must be readable by domain experts such as lawyers and accountants.
It must have precise, unambiguous, and executable semantics.
The respective publishers can use well-established software engineering techniques, such as version control, automated testing, and formal verification.
Checking and ensuring compliance with a new regulation can be as automated as software updates.

We are not the first to pursue this line of thinking~\cite[e.g.,][]{oecd-cracking-the-code,ec-mrer,regelspraak,french-tax-code,catala,blawx}.
Influenced by our background in the professional services business, however, we offer a new spin.
We consider it unrealistic for individuals to become experts in both a legal domain and in software engineering.
Instead, we propose a pair-programming approach: one engineer writing code and one domain expert providing requirements and verifying the implementation.
Accordingly, one key criterion is whether the code is auditable by domain experts, not whether they can write functioning code.
Not only do we think this approach is more realistic, but it also allows us to design an \emph{embedded} language for regulations, making the development of such systems significantly cheaper and faster.

We have designed and implemented \eiger/, a Haskell framework and domain-specific language (DSL) for encoding legal regulations.
\Eiger/ allows a heterogeneous database of records with fields whose values can depend on each other.
Rules for the computation of those values can be written in any order and are written in a straightforward, pure functional style, familiar to users of Excel.
When working with a dataset, \eiger/ accepts inputs with missing information or local overrides of fields that could otherwise be calculated; it allows users to query what extra information must be entered in order to compute a particular result.
\Eiger/ also supports reading and writing JSON, enabling integrations with interactive applications and visualizations, and data pipelines.

This paper presents the following contributions:

\begin{itemize}
\item The user-facing design of \eiger/, designed for a professional services life-cycle process (\S\ref{sec:workflow}), and including the rule-writing interface (\S\ref{sec:writing-rules}), and the end-user interface (\S\ref{sec:ux}).
\item A limited user study (\S\ref{sec:user-study}), where consultants not involved in the development of \eiger/ pair with the \eiger/ \ifanon author(s)\else authors\fi{} to help encode actual regulations.
\item A tour of the implementation of \eiger/ (\S\ref{sec:tour}), involving heavy use of type reflection~\cite{typerep}, generic programming~\cite{generic-deriving,generic-lens-paper,generics-sop-paper}, GHC~9.2's new dot-notation for record access (\ext{Over\-load\-ed\-Record\-Dot}) \cite{overloaded-record-dot-proposal}, and a novel approach to instance retrieval at run-time, all built around a core inspired by \citet{build-systems-a-la-carte}.
\item The \eiger/ implementation itself.\footnote{\url{https://github.com/goldfirere/eiger} \rae{make that exist!}}
\end{itemize}

We conclude with related (\S\ref{sec:related}) and future (\S\ref{sec:future}) work.

\section{Motivating use case: the BEPS regulation}
\label{sec:motivation}
The OECD's Base Erosion and Profit Shifting (BEPS) package addresses tax avoidance by large multinational enterprises.
It ``represents the first substantial renovation of the international tax rules in almost a century'' \cite{beps-model-rules}.

This is a sizable body of regulation with significant impact world-wide.
It is currently one of the major business cases of the professional services industry within the tax domain.
It is also our driving use case for the design of \eiger/ and the accompanying pair-programing methodology.

In the subsequent sections, we take a closer look at BEPS, explain how the industry manages to adopt it today, and highlight some of the benefits of implementation as code.

\subsection{In short: it's complicated}
The OECD has published model rules in a 10-chapter, 70-page document of dense legal prose.
Countries are expected to implement their interpretation of these rules via their respective domestic law.
Implementations will vary because each jurisdiction has different judicial interpretations, tax codes, accounting standards, and monetary policies.
Jurisdictions will put BEPS into force at differing times, and thus there must be transition rules affecting multinational enterprises.

The model rules relate to existing standards such as the International Financial Accounting Standards.
They also introduce new concepts such as the GloBE income, the minimum tax rate, the effective tax rate, the top-up tax, the income inclusion rule, the under-taxed payments rule, and many others.
It offers dozens of elections where reporting entities can choose certain classifications that affect the rules and the tax amount owed.
And it provides an interpretation of ownership structures of multinational organizations, involving concepts such as ultimate parent entities, main entities, permanent establishments, flow-through entities, and constituent entities.
Finally, there are multiple exemptions, exceptions, and amendments for what does and does not count towards the tax owed, based on legal properties of entities, thresholds, prior fiscal years, etc.

\subsection{How we manage today}
\label{sec:how-we-manage-today}
Whenever regulators publish new pieces of draft regulations, professional services go through the following life-cycle process:
\begin{enumerate}
\item \emph{Investment}: Become experts on the regulation.
This involves reading, interpreting, assessing, aligning, and relating, as well as participating in the regulator's Q\&A process and expert discussion groups.
\item \emph{Knowledge management}: Capture the outcome of Phase 1 in a way that allows the business to scale up the number and size of clients.
\item \emph{Delivery} (repeated): Use the outcome of Phase 2 and apply it on client engagements.
\end{enumerate}

Executing this is very challenging in practice.
BEPS has a very broad and interdisciplinary scope.
This has been a trend in regulations for the last decade, as today's problems require increasingly interconnected solutions.
Consequently, experts from many different tax and accounting domains need to come together to align their terminology and establish their interfaces with each other.

Furthermore, while the OECD and governments are incrementally publishing details about the BEPS model rules and their implementation in respective domestic law, regulated entities need to take action today:
parts of BEPS are planned to come into force as early as 2023 in some jurisdictions.
Therefore, the above process is being executed multiple times incrementally, and often in parallel.

Already without the above complications, the knowledge management step can be fragile, since it is based on more prose captured in secondary documents.
This propagates the problems with ambiguity, inconsistency, and incompleteness inherent in this manual, natural-language process.

\subsection{Essential versus accidental complexity}
Software engineers learn how to discern essential from accidental complexity~\cite{silver-bullet}.
Thus, as software engineers, we make the following observation about the problem described in the previous section:
The essential complexity of the problem lies in the semantics of the regulation, trying to capture, supervise, and influence behavior in the real world.
Understanding the concepts, standards and practices involved requires training; grasping their boundaries and nuances requires experience.
It is that knowledge that domain experts offer through professional services, which is where they bring value to clients.

In contrast, the lack of formal specifications of the regulation, the lack of standardized data models and data formats, and the lack of automation in the process is accidental complexity that can and should be removed, along with the corresponding costs for implementation and operation.
A specification in \eiger/ is unambiguous and easy to read for domain experts.
Standard IDE features and related tools make the regulation navigable and discoverable.
Augmented with a comprehensive test suite domain experts can also interact with the regulation through precise and correct examples.
Interactive tools allow for exploration through experimentation and simulation of what-if scenarios, deepening the understanding of the regulation.

Version control can be used to track changes over time and manage adaptions for individual clients via branches.
Version control also allows domain experts to share and evaluate hypothetical or factual changes through merge requests.
Merge requests allow for the implementation of approval processes with controls an audit trail.
Automated type checking and related static analysis yields high confidence that any changes are consistent, while the accompanying changes to the test suite convey their impact.
A continuous delivery pipeline can bring factual changes to all clients in a timely fashion.

These are just a few highlights.
Let's have a closer look how \eiger/ improves the whole professional services life-cycle.

\section{The \eiger/-based life-cycle}
\label{sec:workflow}
It is important to note that \eiger/ addresses two types of users
\begin{enumerate}
\item \emph{Authors}: this is an engineer and domain expert pair that writes the \eiger/ code.
\item \emph{Consultants}: this is the staff on client engagements using an existing \eiger/-based system for delivery.
\end{enumerate}

With that, let's revisit the life-cycle outlined in section \S\ref{sec:how-we-manage-today}, in the context of BEPS, but this time with \eiger/ in mind.
We will explore its benefits and its limitations.

\subsection{Investment and Knowledge Management}
Unless authorities start publishing BEPS and the related domestic law as \eiger/ code, the majority of the investment phase stays the same.
Our limited user study (\S\ref{sec:user-study}), however, suggests that it is beneficial to start capturing insights as \eiger/ code early.
This is even more the case when turning example cases -- used to probe aspects of the regulation -- into executable tests.
Doing so enforces a structure that leads domain experts to a deeper and more precise understanding and guides them towards asking the right questions to eliminate ambiguities, inconsistencies, and gaps.

As the authors unveil the regulation and implement it in \eiger/, they are establishing the source of truth for consultants to draw from.
Together with a comprehensive test suite, this forms the bedrock of knowledge management for this regulation.
The terminology used in the code becomes the lingua franca for the interdisciplinary team.
It is also natural for version-controlled software to evolve iteratively while managing compatibility between versions.
With this, processing the regulation and domestic laws iteratively while delivering on client engagements becomes manageable.

\subsection{Delivery}
\label{sec:eiger-delivery}
A client-specific implementation of BEPS consists of three parts:
\begin{enumerate}
\item \emph{Data Sourcing}: sourcing data from the client's IT systems and mapping them to input data for BEPS.
\item \emph{Application}: applying the BEPS rules to the input data.
\item \emph{Reporting}: rendering the results in a form proper for consumption by stakeholders.
\end{enumerate}

Data sourcing is a significant challenge in practice.
Solving it requires an extensive survey of the client's IT landscape, the data that are available, along with their semantics and quality.
Then, proper mappings between what is available to what is required need to be developed and implemented.

\eiger/ cannot help with most of these problems.
But it can be used to implement the client-specific mapping rules on top of the client-agnostic BEPS rules.
This works well since regulatory rules are, generally speaking, mapping rules and since \eiger/ code is extensible (as demonstrated in \pref{sec:writing-rules}).
With this, the solution for the client can be managed within one holistic framework, with the exception of reporting, which is out of scope for  \eiger/.

\Pref{sec:ux} will further explore how consultants use \eiger/ for delivery and what benefits this brings.
Before that, however, we need a basic understanding of what \eiger/ rules look like.

\section{Writing rules with \eiger/}
\label{sec:writing-rules}

This section demonstrates the \eiger/ framework by looking at parts our implementation of BEPS with \eiger/.

\Eiger/ targets regulations that determine the merits of a case by applying rules that relate to facts.
Example facts include monetary values, Boolean conditions, percentages, etc., but also domain-specific classifications such as the type of a legal entity.
Each fact is either provided as an input value or there is a rule that explains how to determine its value based on the values of other facts.
With this, facts and rules form a directed acyclic graph.

Facts are grouped based on their context.
For example, the context of a fact about income is a concrete legal entity and a concrete fiscal year.
In other words, the basic structure of a regulation implemented in \eiger/ is an entity-relationship model, where facts are properties and entities provide the context for facts.

Let's have a look at the BEPS prototype:
\begin{hscode}\SaveRestoreHook
\column{B}{@{}>{\hspre}l<{\hspost}@{}}%
\column{3}{@{}>{\hspre}l<{\hspost}@{}}%
\column{5}{@{}>{\hspre}l<{\hspost}@{}}%
\column{7}{@{}>{\hspre}l<{\hspost}@{}}%
\column{E}{@{}>{\hspre}l<{\hspost}@{}}%
\>[B]{}\keyword{data}\;\conid{Jurisdiction}\mathrel{=}\conid{MkJ}\;\{\mskip1.5mu {}\<[E]%
\\
\>[B]{}\hsindent{5}{}\<[5]%
\>[5]{}\varid{key}\mathbin{::}\conid{Key}\;\conid{Jurisdiction},{}\<[E]%
\\
\>[B]{}\hsindent{5}{}\<[5]%
\>[5]{}\varid{fiscal\char95 year}\mathbin{::}\conid{Fact}\;\conid{Input}\;\conid{Int},{}\<[E]%
\\[\blanklineskip]%
\>[B]{}\hsindent{5}{}\<[5]%
\>[5]{}\varid{additional\char95 current\char95 top\char95 up\char95 tax},\varid{adjusted\char95 covered\char95 taxes},{}\<[E]%
\\
\>[5]{}\hsindent{2}{}\<[7]%
\>[7]{}\varid{substance\char95 based\char95 income\char95 exclusion}\mathbin{::}\conid{Fact}\;\conid{Optional}\;\conid{Euros},{}\<[E]%
\\
\>[B]{}\hsindent{5}{}\<[5]%
\>[5]{}\varid{top\char95 up\char95 tax\char95 percentage}\mathbin{::}\conid{Fact}\;\conid{Optional}\;\conid{Percentage},\ldots\mskip1.5mu\}{}\<[E]%
\\
\>[B]{}\hsindent{3}{}\<[3]%
\>[3]{}\keyword{deriving}\;\conid{IsRecord}{}\<[E]%
\ColumnHook
\end{hscode}\resethooks

Every record in \eiger/ must have a \ensuremath{\varid{key}} field, used in storing \eiger/ records in a database.
The declaration above additionally says that a jurisdiction belongs to exactly one fiscal year and has a set of facts, such as \ensuremath{\varid{additional\char95 current\char95 top\char95 up\char95 tax}}.
Values for these facts may or may not be present in any given input data set, as indicated by \ensuremath{\conid{Optional}}.
Note that some facts are denominated in \ensuremath{\conid{Euros}}, while \varid{top\_up\_tax\_per\-cen\-tage} is a \ensuremath{\conid{Percentage}}; this separation of types provides an important additional check on the correctness of our implementation of the BEPS rules.

A jurisdiction covers a range of entities:
\begin{hscode}\SaveRestoreHook
\column{B}{@{}>{\hspre}l<{\hspost}@{}}%
\column{3}{@{}>{\hspre}l<{\hspost}@{}}%
\column{5}{@{}>{\hspre}l<{\hspost}@{}}%
\column{7}{@{}>{\hspre}l<{\hspost}@{}}%
\column{9}{@{}>{\hspre}l<{\hspost}@{}}%
\column{20}{@{}>{\hspre}l<{\hspost}@{}}%
\column{40}{@{}>{\hspre}l<{\hspost}@{}}%
\column{60}{@{}>{\hspre}l<{\hspost}@{}}%
\column{E}{@{}>{\hspre}l<{\hspost}@{}}%
\>[B]{}\keyword{data}\;\conid{Entity}\mathrel{=}\conid{MkE}\;\{\mskip1.5mu {}\<[E]%
\\
\>[B]{}\hsindent{5}{}\<[5]%
\>[5]{}\varid{key}\mathbin{::}\conid{Key}\;\conid{Entity},{}\<[E]%
\\
\>[B]{}\hsindent{5}{}\<[5]%
\>[5]{}\varid{fiscal\char95 year}\mathbin{::}\conid{Fact}\;\conid{Input}\;\conid{Int},{}\<[E]%
\\
\>[B]{}\hsindent{5}{}\<[5]%
\>[5]{}\varid{jurisdiction}\mathbin{::}\conid{Fact}\;\conid{Input}\;(\conid{Key}\;\conid{Jurisdiction}),{}\<[E]%
\\[\blanklineskip]%
\>[B]{}\hsindent{5}{}\<[5]%
\>[5]{}\varid{above\char95 the\char95 line\char95 taxes},\varid{adjusted\char95 covered\char95 taxes},{}\<[E]%
\\
\>[5]{}\hsindent{2}{}\<[7]%
\>[7]{}\varid{stock\char95 based\char95 compensation},{}\<[E]%
\\
\>[5]{}\hsindent{2}{}\<[7]%
\>[7]{}\varid{stock\char95 based\char95 compensation\char95 expense},{}\<[E]%
\\
\>[5]{}\hsindent{2}{}\<[7]%
\>[7]{}\varid{stock\char95 based\char95 compensation\char95 deduction},{}\<[E]%
\\
\>[5]{}\hsindent{2}{}\<[7]%
\>[7]{}\varid{stock\char95 based\char95 compensation\char95 expense\char95 expired},{}\<[E]%
\\
\>[5]{}\hsindent{2}{}\<[7]%
\>[7]{}\varid{stock\char95 based\char95 compensation\char95 deduction\char95 expired}{}\<[E]%
\\
\>[7]{}\hsindent{2}{}\<[9]%
\>[9]{}\mathbin{::}\conid{Fact}\;\conid{Optional}\;\conid{Euros},{}\<[E]%
\\
\>[B]{}\hsindent{5}{}\<[5]%
\>[5]{}\varid{stock\char95 based\char95 compensation\char95 election}\mathbin{::}\conid{Fact}\;\conid{Optional}\;\conid{Bool},{}\<[E]%
\\
\>[B]{}\hsindent{5}{}\<[5]%
\>[5]{}\varid{entity\char95 type}\mathbin{::}\conid{Fact}\;\conid{Optional}\;\conid{EntityType},\ldots\mskip1.5mu\}{}\<[E]%
\\
\>[B]{}\hsindent{3}{}\<[3]%
\>[3]{}\keyword{deriving}\;\conid{IsRecord}{}\<[E]%
\\[\blanklineskip]%
\>[B]{}\keyword{data}\;\conid{EntityType}\mathrel{=}{}\<[20]%
\>[20]{}\conid{InvestmentEntity}\mid {}\<[40]%
\>[40]{}\conid{NonSpecialEntity}\mid {}\<[60]%
\>[60]{}\ldots{}\<[E]%
\ColumnHook
\end{hscode}\resethooks

We see that an \ensuremath{\conid{Entity}} contains facts of a few new types, \ensuremath{\conid{Bool}}, and \ensuremath{\conid{EntityType}}.

Let's now have a look at a rule:
\begin{hscode}\SaveRestoreHook
\column{B}{@{}>{\hspre}l<{\hspost}@{}}%
\column{3}{@{}>{\hspre}l<{\hspost}@{}}%
\column{7}{@{}>{\hspre}l<{\hspost}@{}}%
\column{10}{@{}>{\hspre}l<{\hspost}@{}}%
\column{E}{@{}>{\hspre}l<{\hspost}@{}}%
\>[B]{}\mbox{\commentbegin  Article 3.2.2  \commentend}{}\<[E]%
\\
\>[B]{}\keyword{instance}\;\conid{Rule}\;\conid{Entity}\;\text{\ttfamily \char34 stock\char95 based\char95 compensation\char34}\;\keyword{where}{}\<[E]%
\\
\>[B]{}\hsindent{3}{}\<[3]%
\>[3]{}\varid{rule}\;\varid{self}\mathrel{=}\conid{HasRule}\mathbin{\$}{}\<[E]%
\\
\>[3]{}\hsindent{4}{}\<[7]%
\>[7]{}\keyword{if}\;\varid{self}\hsdot{.}{.\,}\varid{stock\char95 based\char95 compensation\char95 election}{}\<[E]%
\\
\>[3]{}\hsindent{4}{}\<[7]%
\>[7]{}\keyword{then}\;\varid{self}\hsdot{.}{.\,}\varid{stock\char95 based\char95 compensation\char95 expense}{}\<[E]%
\\
\>[7]{}\hsindent{3}{}\<[10]%
\>[10]{}\mathbin{-}\varid{self}\hsdot{.}{.\,}\varid{stock\char95 based\char95 compensation\char95 deduction}{}\<[E]%
\\
\>[7]{}\hsindent{3}{}\<[10]%
\>[10]{}\mathbin{+}\varid{self}\hsdot{.}{.\,}\varid{stock\char95 based\char95 compensation\char95 expense\char95 expired}{}\<[E]%
\\
\>[7]{}\hsindent{3}{}\<[10]%
\>[10]{}\mathbin{-}\varid{self}\hsdot{.}{.\,}\varid{stock\char95 based\char95 compensation\char95 deduction\char95 expired}{}\<[E]%
\\
\>[3]{}\hsindent{4}{}\<[7]%
\>[7]{}\keyword{else}\;\mathrm{0}{}\<[E]%
\ColumnHook
\end{hscode}\resethooks

This is a rule for the fact \ensuremath{\varid{stock\char95 based\char95 compensation}} of an \ensuremath{\conid{Entity}}.
This is an example of an Excel-like formula.
We use informal comments to relate code to articles of the legal prose.
The identifier \ensuremath{\varid{self}} is used to access other facts from the same entity.
The rule reads such that if an election of stock-based compensation is made, certain facts are aggregated to yield the stock-based compensation.
If the election is not made, then there is no stock-based compensation.

More advanced rules involve aggregating over a dynamic set of facts.
Here is an example:
\begin{hscode}\SaveRestoreHook
\column{B}{@{}>{\hspre}l<{\hspost}@{}}%
\column{3}{@{}>{\hspre}l<{\hspost}@{}}%
\column{5}{@{}>{\hspre}l<{\hspost}@{}}%
\column{7}{@{}>{\hspre}l<{\hspost}@{}}%
\column{9}{@{}>{\hspre}l<{\hspost}@{}}%
\column{11}{@{}>{\hspre}l<{\hspost}@{}}%
\column{E}{@{}>{\hspre}l<{\hspost}@{}}%
\>[B]{}\mbox{\commentbegin  Article 5.1.1  \commentend}{}\<[E]%
\\
\>[B]{}\keyword{instance}\;\conid{Rule}\;\conid{Jurisdiction}{}\<[E]%
\\
\>[B]{}\hsindent{3}{}\<[3]%
\>[3]{}\text{\ttfamily \char34 adjusted\char95 covered\char95 taxes\char34}\;\keyword{where}{}\<[E]%
\\
\>[3]{}\hsindent{2}{}\<[5]%
\>[5]{}\varid{rule}\;\varid{self}\mathrel{=}\conid{HasRule}\mathbin{\$}{}\<[E]%
\\
\>[5]{}\hsindent{4}{}\<[9]%
\>[9]{}\varid{getAll}\;@\conid{Entity}{}\<[E]%
\\
\>[9]{}\hsindent{2}{}\<[11]%
\>[11]{}\mathbin{`\varid{filteredBy}`}\varid{filter}{}\<[E]%
\\
\>[9]{}\hsindent{2}{}\<[11]%
\>[11]{}\mathbin{`\varid{select}`}(\hsdot{.}{.\,}\varid{adjusted\char95 covered\char95 taxes}){}\<[E]%
\\
\>[9]{}\hsindent{2}{}\<[11]%
\>[11]{}\mathbin{`\varid{andThen}`}\varid{sum}{}\<[E]%
\\
\>[5]{}\hsindent{2}{}\<[7]%
\>[7]{}\keyword{where}{}\<[E]%
\\
\>[7]{}\hsindent{2}{}\<[9]%
\>[9]{}\varid{filter}\;\varid{entity}\mathrel{=}{}\<[E]%
\\
\>[9]{}\hsindent{2}{}\<[11]%
\>[11]{}(\varid{entity}\hsdot{.}{.\,}\varid{jurisdiction}\mathop{==}\varid{self}\hsdot{.}{.\,}\varid{key})\mathbin{`\varid{and}`}{}\<[E]%
\\
\>[9]{}\hsindent{2}{}\<[11]%
\>[11]{}(\varid{entity}\hsdot{.}{.\,}\varid{fiscal\char95 year}\mathop{==}\varid{self}\hsdot{.}{.\,}\varid{fiscal\char95 year})\mathbin{`\varid{and}`}{}\<[E]%
\\
\>[9]{}\hsindent{2}{}\<[11]%
\>[11]{}\mbox{\commentbegin  Article 5.1.3  \commentend}{}\<[E]%
\\
\>[9]{}\hsindent{2}{}\<[11]%
\>[11]{}(\varid{entity}\hsdot{.}{.\,}\varid{entity\char95 type}\mathop{{/}{=}}\varid{use}\;\conid{InvestmentEntity}){}\<[E]%
\ColumnHook
\end{hscode}\resethooks
This rule, resembling an SQL formula, is for the \varid{ad\-jus\-ted\_co\-vered\_tax\-es} fact of a \ensuremath{\conid{Jurisdiction}}.
The rule says that the adjusted covered taxes of a jurisdiction is the sum of the adjusted covered taxes of all entities that meet the filter criteria: the entity must be in that jurisdiction, be associated with the same fiscal year, and not be an \ensuremath{\conid{InvestmentEntity}}.

Finally, some facts have no rules at all:
\begin{hscode}\SaveRestoreHook
\column{B}{@{}>{\hspre}l<{\hspost}@{}}%
\column{3}{@{}>{\hspre}l<{\hspost}@{}}%
\column{5}{@{}>{\hspre}l<{\hspost}@{}}%
\column{E}{@{}>{\hspre}l<{\hspost}@{}}%
\>[B]{}\mbox{\commentbegin  Article 5.4.1  \commentend}{}\<[E]%
\\
\>[B]{}\keyword{instance}\;\conid{Rule}\;\conid{Jurisdiction}{}\<[E]%
\\
\>[B]{}\hsindent{5}{}\<[5]%
\>[5]{}\text{\ttfamily \char34 additional\char95 current\char95 top\char95 up\char95 tax\char34}\;\keyword{where}{}\<[E]%
\\
\>[B]{}\hsindent{3}{}\<[3]%
\>[3]{}\varid{rule}\;\varid{self}\mathrel{=}\conid{NoRule}{}\<[E]%
\ColumnHook
\end{hscode}\resethooks
Facts like \ensuremath{\varid{additional\char95 current\char95 top\char95 up\char95 tax}} are listed as \ensuremath{\conid{Optional}} in order to allow working with a dataset missing that datum, but we still have no way of computing it if it is missing.

If we wanted, we could later decide to introduce a rule for additional current top-up tax that determines the value from raw data.
The rest of the regulation and its implementation would remain unaffected by this extension.
Referring back to \pref{sec:eiger-delivery}, this is the interface between the client-agnostic BEPS core and the client-specific mapping.

Now that we understand what \eiger/ code looks like, we can further explore its use and benefits.

\section{Running your regulations}
\label{sec:ux}
Given the implementation of BEPS with \eiger/ described in the previous section, what does this prove useful during delivery?

Typically, regulated entities adopt regulations in two distinct phases.
\begin{enumerate}
\item \emph{Exploratory:}
This is about understanding the opportunities, benefits, liabilities, and risks that come with the regulation.
There is a strong connection of such an analysis with the entity's business strategy and its business model.
The general question is: how the entity should best position itself for the time when the regulation will be in force?
\item \emph{Operational:}
This stage is about managing compliance with the regulation on an ongoing basis: controls, reporting, and investigations.
At this point it is all about reliability and costs.
\end{enumerate}

The following sections explore how \eiger/ brings benefits to users in both phases.

\subsection{Exploratory}
In the context of BEPS, typical questions in the exploratory phase include:
\begin{itemize}
\item How much additional tax are we projected to owe in each of the next five years?
\item What are the options available to reduce our tax footprint?
\item How much would implementing such a change cost and how much would it save?
\end{itemize}

Exploratory analysis might suggest a wide range of changes to an organization's structure.
For example, it might be recommended to relocate certain entities, acquire, sell, split, or merge other entities, move staff between entities, as well as change the flow of goods and services through multinational subsidiaries.
For each of these ideas, the consultants need to collect relevant input data, such as the entity's financial statements, and apply the rules of the regulation.
Typically a different subset of the rules comes into play each time, so the work tends to not be repeatable.
The potential savings are often in the millions, and mistakes can lead to unpleasant surprises such as heavy fines when interacting with the authorities later.

Today, such work is mostly done manually with light support from Excel sheets and related extract-transform-load and process automation tools.
None of these tools provide the flexibility required to exhaust the available options efficiently and with precision.

An interactive system loaded with a BEPS implementation in \eiger/, in contrast, offers a rich user journey.
Users can browse the data model and the fact graph.
They can also work interactively by providing a set of input data and have the system return a superset which includes values for all facts that can be computed from the provided input.

With this, users can run:
\begin{itemize}
\item \emph{Simple queries}:
The user can ask a regulation-as-code system about the value of any of its facts.
As analysts learn more, they can tailor their queries accordingly.
\item \emph{What-if scenarios}:
Sometimes the user does not yet have the data to determine if a multinational enterprise is eligible, but they still want to know what eligibility would imply.
In such a situation, the user can provide an input value of \ensuremath{\conid{True}} for the fact about eligibility and query the value for the top-up taxes owed.
\item \emph{Optimizations}:
BEPS allows for several elections on how legal entities wish to be classified.
Those elections have an impact on the tax owed, so it is natural to ask which election choices are most preferable.
The user can apply all combinations of elections in turn, and query the tax owed in each case.
\end{itemize}

Based on our experience, none of the tools used in such client engagements today provides this degree of flexibility with high precision and short lead times.

\subsection{Operational}
Once the regulation is in force, certain obligations, such as reporting, have to be met.
Today, this involves extensive IT projects involving consultants, business analysts, developers, testers, and operators.
While this process also yields a machine-executable implementation of the regulation, the resulting code is not auditable by domain experts.
This leaves a gap with plenty of opportunity for misunderstandings between business analysts, developers, and testers.
The implementation is also custom-made, so its costs cannot be mutualized across multiple entities, and the interpretation of the regulation is likely to be unique.
This leads to poor data quality as observed by the authorities across the regulated industry.

Most of these problems are addressed by \eiger/.
A single implementation can serve as the core for the production implementations of all regulated entities.
It is the same well-tested implementation that has been used during the exploratory phase.
For each regulated entity, there is a one-time effort to integrate the \eiger/ core in the respective IT landscape.
It is then fed with raw data and, in contrast to the exploratory phase, regularly executes the same queries using different data once for each reporting period.

\section{User Study}
\label{sec:user-study}
To validate the design of \eiger/ and the approach of pairing domain experts with software engineers, we have run a limited user study.
We paired with a domain expert from each of three different domains and implemented a non-trivial subset of a regulation.
Afterwards, we asked the domain experts the following questions:

\begin{enumerate}
\item How easy was it to get comfortable with the pair-prog\-ram\-ming approach?
\item What's the one thing we could add to make your experience better?
\item How high is your confidence that what you are seeing is a faithful implementation of the legal text?
\item How confident are you that you could read other people's implementation of legal text based on this approach?
\item If, in future, you were asked to advise on regulatory matters, would you use the code or the legal text as a reference?  If not the code, what is missing for you to use this instead?
\item Given the prospective benefits of rules as code as an idea, do you consider the pair programming approach a worthwhile investment?
\item Based on your understanding of the code, are there any specific areas where you think the code would struggle to replicate existing regulatory provisions?
\item What are your main feedback points about the process (both good and bad)?
\end{enumerate}

\Pref{sec:motivation} above provides an introduction to BEPS.
This is the first case of our user study.
As part of this study, and in preparation for a client engagement, we have implemented most of chapters 3, 4, and 5.
The other subjects of study are the EU's Carbon Border Adjustment Mechanism (CBAM) and the UK's Value Added Tax Regime (VAT).
CBAM requires businesses to offset financially the carbon emissions embedded in the products they import from non-EU countries, while VAT is applicable to the sales of goods and services within the UK.
Both choices of subject pushed the design of \eiger/ forward, requiring different features from those used in the BEPS domain.

We implemented a prototype for each case by pairing one of the authors of this paper as an engineer with one of three domain experts.
While the verbatim feedback is available \ifanon in our supplementary material\else in the appendix\fi, we present a summary here.
All three domain experts report feeling comfortable with the pair-programming approach and expressed confidence that the respective implementation is an accurate representation of the respective regulation.
The experts also responded confidently about their ability to read other people's \eiger/ code after a short introduction, eventually using the code as a reference moving forward.
Some experts noted that being forced through the structured process of codifying the regulation deepened their own understanding.
They also expressed surprise how quickly the pair managed to produce functional code and how mutually beneficial it is to work as a pair.
Overall, the experts deemed the investment in this new approach worthwhile, and they considered the approach to be scalable.

From the perspective of the engineer, pair-programming feels very productive.
We can guide the expert with directed questions to help us understand the overall structure first, and then incrementally dive into the details.
As we stumble upon open questions during the implementation, we get immediate answers, keeping the train of thought going.
Sometimes, we need to ask for patience when refactoring or fixing compiler errors.
And, sometimes, we need to take notes to work independently later, when some deeper thinking, research, or a larger cleanup is needed.
So far, it has always been easy to loop the expert in again afterwards.
Finally, it takes some additional effort to keep experimenting with code formatting and implementation techniques to optimize for readability.
To do that well, engineers need a solid understanding of \eiger/, which is provided in the next section.

\section{A Tour of \eiger/}
\label{sec:tour}

There are two key design challenges in implementing \eiger/: designing a mechanism for
writing computation rules for fields in terms of other fields, and making the syntax
for writing these rules straightforward enough for a regulations expert to understand.
Because we want fields of \eiger/ records to be typed (preventing, say, the addition of a monetary amount with a percentage), we must build out some infrastructure to deal with a heterogeneous database (that is, storing many different types of \eiger/ record) with a variety of field types.
Then, despite the complexity under the hood, we must produce an API that is easy for domain experts to read and understand.

\subsection{Tracking dependencies}

An \eiger/ rule is specified by giving a \ensuremath{\conid{ComputationRule}\;\varid{t}}, which is a description of how to compute a value of type \ensuremath{\varid{t}}.
\ensuremath{\conid{ComputationRule}} has a simple definition:
\begin{hscode}\SaveRestoreHook
\column{B}{@{}>{\hspre}l<{\hspost}@{}}%
\column{3}{@{}>{\hspre}l<{\hspost}@{}}%
\column{E}{@{}>{\hspre}l<{\hspost}@{}}%
\>[B]{}\keyword{newtype}\;\conid{ComputationRule}\;\varid{t}\mathrel{=}{}\<[E]%
\\
\>[B]{}\hsindent{3}{}\<[3]%
\>[3]{}\conid{MkRule}\;\{\mskip1.5mu \varid{runComputationRule}\mathbin{::}\conid{EigerM}\;\varid{t}\mskip1.5mu\}{}\<[E]%
\ColumnHook
\end{hscode}\resethooks
A description of how to accomplish a (effectful) task is just a monadic action.
We use our \ensuremath{\conid{EigerM}} monad, whose details are in \pref{sec:EigerM}.
Despite this simplicity, we keep \conid{Comp\-u\-ta\-tion\-Rule} as an abstraction layer for our users, allowing us to change design later.
For example, we might imagine constructing syntax trees describing the operations needed for the rule for further analysis.
Furthermore, \ensuremath{\conid{ComputationRule}} has a number of instances written for it to make it work in our DSL, but we do not want these instances affecting our internal \ensuremath{\conid{EigerM}}.

Within the computations stored in a \ensuremath{\conid{ComputationRule}}, access to data is through two functions: \ensuremath{\varid{accessField}} and
\varid{ac\-cess\-All\-Keys\-Of\-Type}.
Both of these track dependencies, as described below.

\subsubsection{\ensuremath{\varid{accessField}}}

Below is the \ensuremath{\varid{accessField}} function:
\begin{hscode}\SaveRestoreHook
\column{B}{@{}>{\hspre}l<{\hspost}@{}}%
\column{3}{@{}>{\hspre}l<{\hspost}@{}}%
\column{5}{@{}>{\hspre}l<{\hspost}@{}}%
\column{E}{@{}>{\hspre}l<{\hspost}@{}}%
\>[B]{}\varid{accessField}\mathbin{::}\keyword{$\forall$} \hsforall \;(\varid{field\char95 name}\mathbin{::}\conid{Symbol})\;\varid{record}\hsdot{.}{.\,}{}\<[E]%
\\
\>[B]{}\hsindent{3}{}\<[3]%
\>[3]{}(\conid{HasFact}\;\varid{field\char95 name}\;\varid{record},\conid{KnownSymbol}\;\varid{field\char95 name})\Rightarrow {}\<[E]%
\\
\>[B]{}\hsindent{3}{}\<[3]%
\>[3]{}\conid{Key}\;\varid{record}\to \conid{EigerM}\;(\conid{FactType}\;\varid{record}\;\varid{field\char95 name}){}\<[E]%
\\
\>[B]{}\varid{accessField}\;\varid{key}\mathrel{=}\keyword{do}{}\<[E]%
\\
\>[B]{}\hsindent{5}{}\<[5]%
\>[5]{}\varid{recordFieldDependency}\;@\varid{field\char95 name}\;\varid{key}{}\<[E]%
\\
\>[B]{}\hsindent{5}{}\<[5]%
\>[5]{}\varid{record}\leftarrow \varid{getRecordFromKey}\;\varid{key}{}\<[E]%
\\
\>[B]{}\hsindent{5}{}\<[5]%
\>[5]{}\varid{getFact}\;@\varid{field\char95 name}\;\varid{record}{}\<[E]%
\ColumnHook
\end{hscode}\resethooks
The \ensuremath{\varid{accessField}} function is polymorphic in a \ensuremath{\varid{field\char95 name}} (recall that \ensuremath{\conid{Symbol}}s are
just compile-time strings) and a \ensuremath{\varid{record}} type;
these are related by the \ensuremath{\conid{HasFact}} class. A constraint \ensuremath{\conid{HasFact}\;\varid{name}\;\varid{record}}
states that the record type \ensuremath{\varid{record}} has a field named \ensuremath{\varid{name}}. The field
will be a \ensuremath{\conid{Fact}} whose payload is \ensuremath{\conid{FactType}\;\varid{record}\;\varid{name}}, the type returned
by \ensuremath{\varid{accessField}}.
\ensuremath{\conid{FactType}} is an indexed type family~\cite{chak1,closed-type-families}, which can be considered a compile-time function, computing on types.
In this case, \ensuremath{\conid{FactType}} looks at the type definition (via the \ensuremath{\conid{\conid{GHC}.Generics}} interface \cite{generic-deriving}) of the record and extracts out the type of the fact of the given name.
The function is also constrained by a \ensuremath{\conid{KnownSymbol}} constraint;
this says only that we know the field's name at run-time. Its name is needed
only for dependency tracking.

The function takes an argument of type \ensuremath{\conid{Key}\;\varid{record}}, which specifies which record to retrieve from the database.
We will see more about how \ensuremath{\conid{Key}}s in \pref{sec:database}
and how \ensuremath{\varid{getFact}} retrieves the data in \pref{sec:getFact}.

Crucially, the \ensuremath{\varid{accessField}} function also notes internally within the \ensuremath{\conid{EigerM}}
monad that a \ensuremath{\conid{ComputationRule}} required a certain field, via \ensuremath{\varid{recordFieldDependency}}.
This is necessary to support a user mode where an analyst can query what dependencies a given field has.

\subsubsection{\ensuremath{\varid{accessAllKeysOfType}}}

The \ensuremath{\varid{accessAllKeysOfType}} function is defined as follows:
\begin{hscode}\SaveRestoreHook
\column{B}{@{}>{\hspre}l<{\hspost}@{}}%
\column{3}{@{}>{\hspre}l<{\hspost}@{}}%
\column{E}{@{}>{\hspre}l<{\hspost}@{}}%
\>[B]{}\varid{accessAllKeysOfType}\mathbin{::}\keyword{$\forall$} \hsforall \;\varid{record}\hsdot{.}{.\,}\conid{Typeable}\;\varid{record}\Rightarrow {}\<[E]%
\\
\>[B]{}\hsindent{3}{}\<[3]%
\>[3]{}\conid{EigerM}\;[\mskip1.5mu \conid{Key}\;\varid{record}\mskip1.5mu]{}\<[E]%
\\
\>[B]{}\varid{accessAllKeysOfType}\mathrel{=}\keyword{do}{}\<[E]%
\\
\>[B]{}\hsindent{3}{}\<[3]%
\>[3]{}\varid{recordTypeDependency}\;@\varid{record}{}\<[E]%
\\
\>[B]{}\hsindent{3}{}\<[3]%
\>[3]{}\varid{getAllKeysOfType}\mathop{{\langle}{\$}{\rangle}}\varid{getDatabase}{}\<[E]%
\ColumnHook
\end{hscode}\resethooks

This function uses a run-time type representation (retrievable from the \ensuremath{\conid{Typeable}\;\varid{record}} constraint) for a record type to retrieve a list of all keys to records of that type.
Once again, the \ensuremath{\varid{accessAllKeysOfType}} function remembers this call, via invoking \ensuremath{\varid{recordTypeDependency}}.
This way, an analyst can learn that, for example, adding a new \ensuremath{\conid{Entity}} with a certain \ensuremath{\conid{Jurisdiction}} might affect the calculation of \ensuremath{\conid{Jurisdiction}} fields.

\subsubsection{The \ensuremath{\conid{EigerM}} monad}
\label{sec:EigerM}

The \ensuremath{\conid{EigerM}} monad is used \\ throughout \eiger/ to structure our computations. It has
a mix of standard monad features. \ensuremath{\conid{EigerM}} is:
\begin{itemize}
\item \ensuremath{\conid{State}}-like over our \ensuremath{\conid{Database}} (\S\ref{sec:database});
\item \ensuremath{\conid{Writer}}-like over a \ensuremath{\conid{Dependencies}} monoid tracking dependencies; and
\item \ensuremath{\conid{Except}}-like over a list of errors, including missing dependencies.
\end{itemize}

One detail in this list may be surprising, in that monadic actions can result in a \emph{list}
of errors. This is unusual: a monad able to throw exceptions often throws only one!

We can explain this twist with an example. Suppose we wish to compute a total income
from an unearned income and an earned income, but we actually have neither fact in our data set.
We might imagine a computation that looks like this (in a more Haskell-like syntax than
our DSL as demonstrated in \pref{sec:writing-rules}):

\begin{hscode}\SaveRestoreHook
\column{B}{@{}>{\hspre}l<{\hspost}@{}}%
\column{3}{@{}>{\hspre}l<{\hspost}@{}}%
\column{13}{@{}>{\hspre}c<{\hspost}@{}}%
\column{13E}{@{}l@{}}%
\column{17}{@{}>{\hspre}l<{\hspost}@{}}%
\column{E}{@{}>{\hspre}l<{\hspost}@{}}%
\>[B]{}\varid{getTotal}\;\varid{compute\char95 unearned}\;\varid{compute\char95 earned}\mathrel{=}\keyword{do}{}\<[E]%
\\
\>[B]{}\hsindent{3}{}\<[3]%
\>[3]{}\varid{unearned}{}\<[13]%
\>[13]{}\leftarrow {}\<[13E]%
\>[17]{}\varid{compute\char95 unearned}{}\<[E]%
\\
\>[B]{}\hsindent{3}{}\<[3]%
\>[3]{}\varid{earned}{}\<[13]%
\>[13]{}\leftarrow {}\<[13E]%
\>[17]{}\varid{compute\char95 earned}{}\<[E]%
\\
\>[B]{}\hsindent{3}{}\<[3]%
\>[3]{}\varid{return}\;(\varid{unearned}\mathbin{+}\varid{earned}){}\<[E]%
\ColumnHook
\end{hscode}\resethooks

However, if there is no way to \ensuremath{\varid{compute\char95 unearned}}, the first line
here will throw an exception. No other possibility exists because
there is no valid answer to return from the computation. Yet, from
a user standpoint, it would be a shame not to learn that \ensuremath{\varid{compute\char95 earned}}
is just as impossible as \ensuremath{\varid{compute\char95 unearned}}.

The \ensuremath{\conid{EigerM}} monad thus has a carefully written \ensuremath{\conid{Applicative}} instance.
In its definition for \ensuremath{(\mathop{{\langle}{*}{\rangle}})\mathbin{::}\conid{Applicative}\;\varid{f}\Rightarrow \varid{f}\;(\varid{a}\to \varid{b})\to \varid{f}\;\varid{a}\to \varid{f}\;\varid{b}},
we run the second action \emph{even if the first one fails}.
This behavior makes our instances not quite lawful: the laws for \ensuremath{\conid{Applicative}}s
state that \ensuremath{\conid{Applicative}}s that are also \ensuremath{\conid{Monad}}s must obey the law saying that
\ensuremath{\varid{m}_{1}\mathop{{\langle}{*}{\rangle}}\varid{m}_{2}} is identical to \ensuremath{\varid{m}_{1}\bind \lambda \varid{x1}\to \varid{m}_{2}\bind \lambda \varid{x2}\to \varid{return}\;(\varid{x1}\;\varid{x2})}.
Yet this is not the case for \ensuremath{\conid{EigerM}}: \ensuremath{\varid{m}_{1}\mathop{{\langle}{*}{\rangle}}\varid{m}_{2}} performs both actions,
while the right-hand side of the law performs only \ensuremath{\varid{m}_{1}} if \ensuremath{\varid{m}_{1}} fails.

It is tempting to try to bring the \ensuremath{\conid{Applicative}} instance and the \ensuremath{\conid{Monad}}
instance back into alignment, but this is impossible. In general, computations
may need \ensuremath{\conid{Monad}}-like behavior: we might have one fact in our database that
tells us where to look next. For example, our database might give us the keys
of records describing the dependent children of a taxpayer; only after retrieving those
keys can we access the incomes of the dependent children. Accordingly, in these
scenarios, our dependency tracking is on a ``best effort'' basis: we can look
for dependencies only with the information we have, and it is possible that
providing more data for a computation will reveal more dependencies. In order
to allow the possibility of conditional computations, this situation is unavoidable.

Is it problematic to have these unlawful instances? In our opinion, no. The
\ensuremath{\conid{Applicative}} combinators will be more informative than their \ensuremath{\conid{Monad}} counterparts,
but both results are guaranteed to be subsets of the true set of dependencies.
Furthermore, it seems unhelpful to users to allow strict compliance with
laws to stop us from reporting as many dependencies as possible.
\rae{Insert good self-referential joke about the value of compliance with regulations here.}

Our situation is rather like that faced by the \ensuremath{\conid{Haxl}} monad described by
\citet[Section 5.4]{haxl}, where their \ensuremath{\conid{Applicative}} instance leads to better performance
than their \ensuremath{\conid{Monad}} instance. Just like them, we think it is acceptable to bend the
laws here.\footnote{The \ensuremath{\conid{Haxl}} monad was the motivating example for the
\ext{ApplicativeDo} extension~\cite{applicative-do}, which desugars \ensuremath{\keyword{do}}-notation
to use \ensuremath{\conid{Applicative}} combinators where possible. However, our
user-facing syntax avoids \ensuremath{\keyword{do}}-notation, so we do not use \ext{ApplicativeDo},
though it would indeed work well for \eiger/ computations written with \ensuremath{\keyword{do}}-notation.}

\subsubsection{User interface}

The design described here supports the following client-facing functions:

\begin{hscode}\SaveRestoreHook
\column{B}{@{}>{\hspre}l<{\hspost}@{}}%
\column{3}{@{}>{\hspre}l<{\hspost}@{}}%
\column{E}{@{}>{\hspre}l<{\hspost}@{}}%
\>[B]{}\varid{get}\mathbin{::}\keyword{$\forall$} \hsforall \;(\varid{field\char95 name}\mathbin{::}\conid{Symbol})\;\varid{record}\hsdot{.}{.\,}{}\<[E]%
\\
\>[B]{}\hsindent{3}{}\<[3]%
\>[3]{}\conid{HasFact}\;\varid{field\char95 name}\;\varid{record}\Rightarrow {}\<[E]%
\\
\>[B]{}\hsindent{3}{}\<[3]%
\>[3]{}\conid{Key}\;\varid{record}\to \conid{EigerM}\;(\conid{FactType}\;\varid{record}\;\varid{field\char95 name}){}\<[E]%
\\
\>[B]{}\varid{getMissingDependencies}\mathbin{::}\keyword{$\forall$} \hsforall \;(\varid{field\char95 name}\mathbin{::}\conid{Symbol})\;\varid{record}\hsdot{.}{.\,}{}\<[E]%
\\
\>[B]{}\hsindent{3}{}\<[3]%
\>[3]{}\conid{HasFact}\;\varid{field\char95 name}\;\varid{record}\Rightarrow \conid{Key}\;\varid{record}\to {}\<[E]%
\\
\>[B]{}\hsindent{3}{}\<[3]%
\>[3]{}\conid{EigerM}\;([\mskip1.5mu \conid{String}\mskip1.5mu],[\mskip1.5mu \conid{TyCon}\mskip1.5mu]){}\<[E]%
\ColumnHook
\end{hscode}\resethooks

The \ensuremath{\varid{get}} function retrieves the value of the field named \ensuremath{\varid{field\char95 name}} from the record keyed by the key provided.
It uses rules as necessary to compute any field that is not present in the database and updates the database as it works.

The \ensuremath{\varid{getMissingDependencies}} function returns lists of all facts that are needed to compute
a target fact (but are not present), as well as a list of all record
types that were queried via \ensuremath{\varid{accessAllKeysOfType}}. The function is useful against
a data set that is missing some input facts. For example, if we want to
know the total income of an individual and know their earned income but
not their unearned income, this function would say that we must input
the unearned income. The list of record types consulted by \ensuremath{\varid{accessAllKeysOfType}}
is important because an addition (or deletion) of such a record might
change the result, as well.

In addition to \ensuremath{\varid{getMissingDependencies}}, we can imagine further queries
of this form, including listing all dependencies (missing or not).

\subsection{A typed interface}

This section explores the challenges of working in a richly typed environment,
and how \eiger/ overcomes these challenges.

\subsubsection{A heterogeneous \ensuremath{\conid{Database}}}
\label{sec:database}

An \eiger/ data set is \emph{heterogeneous}: it contains records of many
different types, such as the \ensuremath{\conid{Jurisdiction}} and \ensuremath{\conid{Entity}} types of \pref{sec:writing-rules}.
We must store these different types in one database. Furthermore, this database
must be designed without knowledge of what types it will be storing, and yet
we wish to avoid extraneous run-time type checking.

The solution is to use a \emph{dependent map} \\ \cite{dependent-map-package}. The \package{dependent-map}
Has\-kell package exports the following type and functions:

\begin{hscode}\SaveRestoreHook
\column{B}{@{}>{\hspre}l<{\hspost}@{}}%
\column{E}{@{}>{\hspre}l<{\hspost}@{}}%
\>[B]{}\keyword{type}\;\conid{DMap}\mathbin{::}\keyword{$\forall$} \hsforall \;\varid{ki}\hsdot{.}{.\,}(\varid{ki}\to \conid{Type})\to (\varid{ki}\to \conid{Type})\to \conid{Type}{}\<[E]%
\\
\>[B]{}\keyword{data}\;\conid{DMap}\;\varid{k}\;\varid{f}\mathrel{=}\ldots{}\<[E]%
\\[\blanklineskip]%
\>[B]{}\varid{lookup}\mathbin{::}\conid{GCompare}\;\varid{k}\Rightarrow \varid{k}\;\varid{v}\to \conid{DMap}\;\varid{k}\;\varid{f}\to \conid{Maybe}\;(\varid{f}\;\varid{v}){}\<[E]%
\\
\>[B]{}\varid{insert}\mathbin{::}\conid{GCompare}\;\varid{k}\Rightarrow \varid{k}\;\varid{v}\to \varid{f}\;\varid{v}\to \conid{DMap}\;\varid{k}\;\varid{f}\to \conid{DMap}\;\varid{k}\;\varid{f}{}\<[E]%
\ColumnHook
\end{hscode}\resethooks

A \ensuremath{\conid{DMap}\;\varid{k}\;\varid{f}} is a finite map connecting keys of type \ensuremath{\varid{k}\;\varid{v}}
to values of type \ensuremath{\varid{f}\;\varid{v}}, for any \ensuremath{\varid{v}}. Accordingly, \ensuremath{\varid{lookup}}
and \ensuremath{\varid{insert}} work over keys \ensuremath{\varid{k}\;\varid{v}} and values \ensuremath{\varid{f}\;\varid{v}}; these
functions are polymorphic over the choice of \ensuremath{\varid{v}}. Using \ensuremath{\conid{DMap}},
we can map a key of type \ensuremath{\conid{Key}\;\varid{record}} to a value of type \ensuremath{\varid{record}},
using the following wrapper, defined as part of \eiger/:

\begin{hscode}\SaveRestoreHook
\column{B}{@{}>{\hspre}l<{\hspost}@{}}%
\column{E}{@{}>{\hspre}l<{\hspost}@{}}%
\>[B]{}\keyword{type}\;\conid{DatabaseValue}\mathbin{::}\conid{Type}\to \conid{Type}{}\<[E]%
\\
\>[B]{}\keyword{data}\;\conid{DatabaseValue}\;\varid{rec}\mathrel{=}\conid{MkDbV}\;(\conid{TypeRep}\;\varid{rec})\;\varid{rec}{}\<[E]%
\ColumnHook
\end{hscode}\resethooks

We see here that a \ensuremath{\conid{DatabaseValue}} stores both a run-time
type representation of a record (this is \ensuremath{\conid{TypeRep}\;\varid{record}}) and
a value of type \ensuremath{\varid{record}}. The \ensuremath{\conid{TypeRep}\;\varid{record}} is necessary to
implement \ensuremath{\varid{getAllKeysOfType}}, as presented in the \ensuremath{\conid{Database}} interface
in \pref{fig:database}.

\begin{figure}
\begin{hscode}\SaveRestoreHook
\column{B}{@{}>{\hspre}l<{\hspost}@{}}%
\column{3}{@{}>{\hspre}l<{\hspost}@{}}%
\column{15}{@{}>{\hspre}c<{\hspost}@{}}%
\column{15E}{@{}l@{}}%
\column{18}{@{}>{\hspre}l<{\hspost}@{}}%
\column{34}{@{}>{\hspre}l<{\hspost}@{}}%
\column{E}{@{}>{\hspre}l<{\hspost}@{}}%
\>[B]{}\keyword{data}\;\conid{Key}\;\varid{record}\mathrel{=}\conid{UnsafeMkKey}\;(\conid{TypeRep}\;\varid{record})\;\conid{UserKey}{}\<[E]%
\\
\>[B]{}\keyword{data}\;\conid{UserKey}{}\<[15]%
\>[15]{}\mathrel{=}{}\<[15E]%
\>[18]{}\conid{ExternalUserKey}\;\conid{Text}{}\<[E]%
\\
\>[15]{}\mid {}\<[15E]%
\>[18]{}\conid{InternalUserKey}\;\conid{InternalKey}{}\<[E]%
\\
\>[B]{}\keyword{newtype}\;\conid{InternalKey}\mathrel{=}\conid{MkIK}\;\conid{Int}{}\<[34]%
\>[34]{}\mbox{\onelinecomment  kept abstract}{}\<[E]%
\\[\blanklineskip]%
\>[B]{}\keyword{data}\;\conid{Database}\mathrel{=}{}\<[E]%
\\
\>[B]{}\hsindent{3}{}\<[3]%
\>[3]{}\conid{MkDb}\;\{\mskip1.5mu \varid{database}\mathbin{::}\conid{DMap}\;\conid{Key}\;\conid{DatabaseValue},\ldots\mskip1.5mu\}{}\<[E]%
\\
\>[B]{}\varid{databaseLookup}\mathbin{::}\conid{Database}\to \conid{Key}\;\varid{record}\to \conid{Maybe}\;\varid{record}{}\<[E]%
\\
\>[B]{}\varid{getAllKeysOfType}\mathbin{::}\conid{Typeable}\;\varid{rec}\Rightarrow \conid{Database}\to [\mskip1.5mu \conid{Key}\;\varid{rec}\mskip1.5mu]{}\<[E]%
\ColumnHook
\end{hscode}\resethooks
\caption{Excerpt of \ensuremath{\conid{Database}} interface}
\label{fig:database}
\end{figure}

We see in \pref{fig:database} also the function \ensuremath{\varid{databaseLookup}}, which
finds the record associated with a key, the critical operation in our
heterogeneous database; the implementation of this function is a straightforward
use of \ensuremath{\varid{lookup}} from the \package{dependent-map} package.

\Pref{fig:database} also presents the \ensuremath{\conid{Key}} type. A \ensuremath{\conid{Key}} stores a
run-time type representation of a record, as well as a unique identifier
for the record to look up. \Eiger/ supports both user-specified identifiers,
stored as \ensuremath{\conid{Text}}, as well as internally generated keys, represented simply
as \ensuremath{\conid{Int}}s. The elided fields of a \ensuremath{\conid{Database}} include some mundane structures that
guarantee uniqueness of \ensuremath{\conid{UserKey}}s in the database.

A \ensuremath{\conid{Key}} needs a type representation in order to safely implement the \ensuremath{\conid{GCompare}}
class, exported in the \package{some} package~\cite{some-package}:

\begin{hscode}\SaveRestoreHook
\column{B}{@{}>{\hspre}l<{\hspost}@{}}%
\column{3}{@{}>{\hspre}l<{\hspost}@{}}%
\column{13}{@{}>{\hspre}c<{\hspost}@{}}%
\column{13E}{@{}l@{}}%
\column{17}{@{}>{\hspre}l<{\hspost}@{}}%
\column{E}{@{}>{\hspre}l<{\hspost}@{}}%
\>[B]{}\keyword{type}\;\conid{GEq}\mathbin{::}\keyword{$\forall$} \hsforall \;\varid{k}\hsdot{.}{.\,}(\varid{k}\to \conid{Type})\to \conid{Constraint}{}\<[E]%
\\
\>[B]{}\keyword{class}\;\conid{GEq}\;\varid{f}\;\keyword{where}\;\varid{geq}\mathbin{::}\varid{f}\;\varid{a}\to \varid{f}\;\varid{b}\to \conid{Maybe}\;(\varid{a}\mathrel{{:}{\sim}{:}}\varid{b}){}\<[E]%
\\
\>[B]{}\keyword{class}\;\conid{GEq}\;\varid{f}\Rightarrow \conid{GCompare}\;\varid{f}\;\keyword{where}{}\<[E]%
\\
\>[B]{}\hsindent{3}{}\<[3]%
\>[3]{}\varid{gcompare}\mathbin{::}\varid{f}\;\varid{a}\to \varid{f}\;\varid{b}\to \conid{GOrdering}\;\varid{a}\;\varid{b}{}\<[E]%
\\
\>[B]{}\keyword{data}\;\conid{GOrdering}\;\varid{a}\;\varid{b}\;\keyword{where}{}\<[E]%
\\
\>[B]{}\hsindent{3}{}\<[3]%
\>[3]{}\conid{GLT},\conid{GGT}{}\<[13]%
\>[13]{}\mathbin{::}{}\<[13E]%
\>[17]{}\conid{GOrdering}\;\varid{a}\;\varid{b}{}\<[E]%
\\
\>[B]{}\hsindent{3}{}\<[3]%
\>[3]{}\conid{GEQ}{}\<[13]%
\>[13]{}\mathbin{::}{}\<[13E]%
\>[17]{}\conid{GOrdering}\;\varid{t}\;\varid{t}{}\<[E]%
\ColumnHook
\end{hscode}\resethooks

Note that the result type of \ensuremath{\varid{gcompare}} is a \ensuremath{\conid{GOrdering}}, whose \ensuremath{\conid{GEQ}}
constructor requires that the indices of the types being compared
(that is, \ensuremath{\varid{a}} and \ensuremath{\varid{b}}) are the same. If a \ensuremath{\conid{Key}} stored only the unique
identifier \ensuremath{\conid{UserKey}}, there would be no way to prove to the type checker
that when two \ensuremath{\conid{UserKey}}s are equal, the types of the records they
index are also equal. The \ensuremath{\conid{TypeRep}} stored in a \ensuremath{\conid{Key}} allows us to make
this check and establish type safety.

\subsubsection{Field access}
\label{sec:has-field}

Suppose we have a record \ensuremath{\varid{r}} of type \ensuremath{\varid{record}}. We will need a way of
extracting the value of a field \ensuremath{\varid{name}}. Recent versions of GHC
provide its \ensuremath{\conid{HasField}} class \cite{has-field-proposal}, which would seem to do the
trick:
\begin{hscode}\SaveRestoreHook
\column{B}{@{}>{\hspre}l<{\hspost}@{}}%
\column{3}{@{}>{\hspre}l<{\hspost}@{}}%
\column{E}{@{}>{\hspre}l<{\hspost}@{}}%
\>[B]{}\keyword{class}\;\conid{HasField}\;\varid{x}\;\varid{r}\;\varid{a}\mid \varid{x}\;\varid{r}\to \varid{a}\;\keyword{where}{}\<[E]%
\\
\>[B]{}\hsindent{3}{}\<[3]%
\>[3]{}\varid{getField}\mathbin{::}\varid{r}\to \varid{a}{}\<[E]%
\ColumnHook
\end{hscode}\resethooks
In fact, \ensuremath{\varid{getField}\;@\varid{name}\;\varid{r}} works. However, this approach is
insufficient for our needs. Not only do we need to access fields,
we need to \emph{update} them after performing a computation on a field
derived from others via user-written rules. GHC has no facility
analogous to \ensuremath{\varid{getField}} for updating fields, due to debates about 
what the best design is.\footnote{There is an accepted proposal~\cite{set-field-proposal},
but it has not been implemented and personal communication suggests a revision
of the plan is in the works.}

The \package{generic-lens} package~\cite{generic-lens-paper} offers a way forward.
The package exports these functions:
\begin{hscode}\SaveRestoreHook
\column{B}{@{}>{\hspre}l<{\hspost}@{}}%
\column{E}{@{}>{\hspre}l<{\hspost}@{}}%
\>[B]{}\varid{getField}\mathbin{::}\keyword{$\forall$} \hsforall \;\varid{f}\;\varid{a}\;\varid{s}\hsdot{.}{.\,}\conid{HasField'}\;\varid{f}\;\varid{s}\;\varid{a}\Rightarrow \varid{s}\to \varid{a}{}\<[E]%
\\
\>[B]{}\varid{setField}\mathbin{::}\keyword{$\forall$} \hsforall \;\varid{f}\;\varid{a}\;\varid{s}\hsdot{.}{.\,}\conid{HasField'}\;\varid{f}\;\varid{s}\;\varid{a}\Rightarrow \varid{a}\to \varid{s}\to \varid{s}{}\<[E]%
\ColumnHook
\end{hscode}\resethooks
The \ensuremath{\conid{HasField'}} class is part of \package{generic-lens}; it is distinct
from the \ensuremath{\conid{HasField}} class above that is part of GHC. \ensuremath{\conid{HasField'}} instances
are built using the generic representations from  \\ \ensuremath{\conid{\conid{GHC}.Generics}}~\cite{generic-deriving}.

Because we wish to use functions named \ensuremath{\varid{getField}} and \ensuremath{\varid{setField}} as part
of the interface presented to authors writing rules (via the \ext{OverloadedRecordDot}
and \ext{RebindableSyntax} extensions; see \pref{sec:overloaded-record-dot}),
we rename and repackage these functions for internal use, where \ensuremath{\conid{GL}} is a
module alias to the appropriate module from \package{generic-lens}:

\begin{hscode}\SaveRestoreHook
\column{B}{@{}>{\hspre}l<{\hspost}@{}}%
\column{3}{@{}>{\hspre}l<{\hspost}@{}}%
\column{E}{@{}>{\hspre}l<{\hspost}@{}}%
\>[B]{}\mbox{\onelinecomment  this includes \ensuremath{\conid{Generic}}, required for a \ensuremath{\conid{HasField'}} instance}{}\<[E]%
\\
\>[B]{}\keyword{type}\;\conid{HasRecordField}\;\varid{field\char95 name}\;\varid{rec}\;\varid{field\char95 ty}\mathrel{=}{}\<[E]%
\\
\>[B]{}\hsindent{3}{}\<[3]%
\>[3]{}(\conid{HasField'}\;\varid{field\char95 name}\;\varid{rec}\;\varid{field\char95 ty},\conid{Generic}\;\varid{rec}){}\<[E]%
\\[\blanklineskip]%
\>[B]{}\varid{getRecordField}\mathbin{::}\keyword{$\forall$} \hsforall \;\varid{name}\;\varid{rec}\;\varid{ty}\hsdot{.}{.\,}{}\<[E]%
\\
\>[B]{}\hsindent{3}{}\<[3]%
\>[3]{}\conid{HasRecordField}\;\varid{name}\;\varid{rec}\;\varid{ty}\Rightarrow \varid{rec}\to \varid{ty}{}\<[E]%
\\
\>[B]{}\varid{getRecordField}\mathrel{=}\varid{\conid{GL}.getField}\;@\varid{name}{}\<[E]%
\\[\blanklineskip]%
\>[B]{}\varid{setRecordField}\mathbin{::}\keyword{$\forall$} \hsforall \;\varid{name}\;\varid{rec}\;\varid{ty}\hsdot{.}{.\,}{}\<[E]%
\\
\>[B]{}\hsindent{3}{}\<[3]%
\>[3]{}\conid{HasRecordField}\;\varid{name}\;\varid{rec}\;\varid{ty}\Rightarrow \varid{rec}\to \varid{ty}\to \varid{rec}{}\<[E]%
\\
\>[B]{}\varid{setRecordField}\;\varid{r}\;\varid{x}\mathrel{=}\varid{\conid{GL}.setField}\;@\varid{name}\;\varid{x}\;\varid{r}{}\<[E]%
\ColumnHook
\end{hscode}\resethooks

\subsubsection{Facts}

As we saw in \pref{sec:writing-rules}, an \eiger/ record contains fields
declaring \ensuremath{\conid{Fact}}s. This type and related definitions are here:

\begin{hscode}\SaveRestoreHook
\column{B}{@{}>{\hspre}l<{\hspost}@{}}%
\column{32}{@{}>{\hspre}c<{\hspost}@{}}%
\column{32E}{@{}l@{}}%
\column{35}{@{}>{\hspre}l<{\hspost}@{}}%
\column{E}{@{}>{\hspre}l<{\hspost}@{}}%
\>[B]{}\keyword{data}\;\conid{FactSort}\mathrel{=}\conid{Input}\mid \conid{Optional}{}\<[E]%
\\[\blanklineskip]%
\>[B]{}\keyword{type}\;\conid{Fact}\mathbin{::}\conid{FactSort}\to \conid{Type}\to \conid{Type}{}\<[E]%
\\
\>[B]{}\keyword{data}\;\keyword{family}\;\conid{Fact}\;\varid{sort}\;\varid{t}{}\<[E]%
\\[\blanklineskip]%
\>[B]{}\keyword{newtype}\;\keyword{instance}\;\conid{Fact}\;\conid{Input}\;\varid{t}\mathrel{=}\conid{InputFact}\;\{\mskip1.5mu \varid{unInputFact}\mathbin{::}\varid{t}\mskip1.5mu\}{}\<[E]%
\\
\>[B]{}\keyword{data}\;\keyword{instance}\;\conid{Fact}\;\conid{Optional}\;\varid{t}{}\<[32]%
\>[32]{}\mathrel{=}{}\<[32E]%
\>[35]{}\conid{MustDerive}{}\<[E]%
\\
\>[32]{}\mid {}\<[32E]%
\>[35]{}\conid{AlreadyKnown}\;\varid{t}{}\<[E]%
\\[\blanklineskip]%
\>[B]{}\keyword{type}\;\conid{GetFieldFactSort}\mathbin{::}\conid{Type}\to \conid{Symbol}\to \conid{FactSort}{}\<[E]%
\\
\>[B]{}\keyword{type}\;\keyword{family}\;\conid{GetFieldFactSort}\;\varid{record}\;\varid{name}{}\<[E]%
\ColumnHook
\end{hscode}\resethooks

We first define a \ensuremath{\conid{FactSort}} as either \ensuremath{\conid{Input}} or \ensuremath{\conid{Optional}}. An \ensuremath{\conid{Input}}
fact must be specified in order for a record to be well formed; an \ensuremath{\conid{Optional}}
fact might be missing. Some \ensuremath{\conid{Optional}} facts have computation rules associated
with them, but others do not -- a fact missing in the input and with no
computation rule is simply unknown.

The \ensuremath{\conid{Fact}} type is a \emph{data family}~\cite{chak2}. Data families
are like the more common type families in that they allow pattern-matching
on their arguments, but a data instance is required to produce a fresh datatype or
a newtype. Accordingly, data family types can appear unsaturated in Haskell
(type families must be applied to all of their arguments), and it is possible to
write type (or data) families that pattern-match on data families.

This ability to match is essential in the implementation of \ensuremath{\conid{GetFieldFactSort}}.
This type family, needed to define \ensuremath{\conid{HasFact}} instances (in the next section),
takes a record type and a field name, returning the \ensuremath{\conid{FactSort}} of the fact
stored in that field. It does this in two steps: first, \ensuremath{\conid{GetFieldFactSort}}
must retrieve the type of the field; then, it must extract out the \ensuremath{\conid{FactSort}}
from something like \ensuremath{\conid{Fact}\;\conid{Input}\;\conid{EuroAmount}}. The second step requires that
\ensuremath{\conid{Fact}} be a \emph{data} family, not a \emph{type} family.

To implement the first step, we must have a way of accessing a field's
type given its enclosing record type and the field's name. Sadly, the \ensuremath{\conid{HasField}}
interface of \citet{has-field-proposal} works via a functional
dependency and is thus inadequate for obtaining a field's type from its
name and enclosing record type. Instead, we must build off
the \ensuremath{\conid{\conid{GHC}.Generics}} library (details of the type families are elided due to space)
to finish the
implementation of \ensuremath{\conid{GetFieldFactSort}}.

\subsubsection{\ensuremath{\conid{HasFact}}}
\label{sec:getFact}

The fact retrieval mechanism centers on the \ensuremath{\conid{HasFact}} constraint.
We cannot, however, define this as a class directly. This is because
we must write separate instances for \ensuremath{\conid{Input}} facts and for \ensuremath{\conid{Optional}}
facts; yet whether a fact is an \ensuremath{\conid{Input}} or is \ensuremath{\conid{Optional}} is not
apparent from the record type and field name. We thus create the internal
\ensuremath{\conid{HasFact\char95 }} class and define \ensuremath{\conid{HasFact}} in terms of \ensuremath{\conid{HasFact\char95 }}:

\begin{hscode}\SaveRestoreHook
\column{B}{@{}>{\hspre}l<{\hspost}@{}}%
\column{3}{@{}>{\hspre}l<{\hspost}@{}}%
\column{5}{@{}>{\hspre}l<{\hspost}@{}}%
\column{6}{@{}>{\hspre}l<{\hspost}@{}}%
\column{7}{@{}>{\hspre}l<{\hspost}@{}}%
\column{9}{@{}>{\hspre}l<{\hspost}@{}}%
\column{11}{@{}>{\hspre}l<{\hspost}@{}}%
\column{13}{@{}>{\hspre}l<{\hspost}@{}}%
\column{165}{@{}>{\hspre}l<{\hspost}@{}}%
\column{E}{@{}>{\hspre}l<{\hspost}@{}}%
\>[B]{}\keyword{type}\;\conid{HasFact\char95 }\mathbin{::}\conid{Symbol}\to \conid{Type}\to \conid{FactSort}\to \conid{Constraint}{}\<[E]%
\\
\>[B]{}\keyword{class}\;(\ldots)\Rightarrow \conid{HasFact\char95 }\;\varid{field\char95 name}\;\varid{record}\;\varid{sort}\;\keyword{where}{}\<[E]%
\\
\>[B]{}\hsindent{3}{}\<[3]%
\>[3]{}\varid{getFact}\mathbin{::}\varid{record}\to \conid{EigerM}\;(\conid{FactType}\;\varid{record}\;\varid{field\char95 name}){}\<[E]%
\\[\blanklineskip]%
\>[B]{}\keyword{type}\;\conid{HasFact}\;\varid{field\char95 name}\;\varid{rec}\mathrel{=}{}\<[E]%
\\
\>[B]{}\hsindent{3}{}\<[3]%
\>[3]{}\conid{HasFact\char95 }\;\varid{field\char95 name}\;\varid{rec}\;(\conid{GetFieldFactSort}\;\varid{rec}\;\varid{field\char95 name}){}\<[E]%
\\[\blanklineskip]%
\>[B]{}\keyword{instance}\;\{-{\#}\;\keyword{overlapping}\;{\#}{-}\}{}\<[E]%
\\
\>[B]{}\hsindent{3}{}\<[3]%
\>[3]{}({}\<[6]%
\>[6]{}\ldots,\conid{FactType}\;\varid{record}\;\text{\ttfamily \char34 key\char34}\,\sim\,\conid{Key}\;\varid{record})\Rightarrow {}\<[E]%
\\
\>[3]{}\hsindent{2}{}\<[5]%
\>[5]{}\conid{HasFact\char95 }\;\text{\ttfamily \char34 key\char34}\;\varid{record}\;\conid{Input}\;\keyword{where}\;\ldots{}\<[E]%
\\[\blanklineskip]%
\>[B]{}\keyword{instance}\;({}\<[13]%
\>[13]{}\ldots,\conid{GetFieldFactSort}\;\varid{record}\;\varid{name}\,\sim\,\conid{Input})\Rightarrow {}\<[E]%
\\
\>[B]{}\hsindent{5}{}\<[5]%
\>[5]{}\conid{HasFact\char95 }\;\varid{name}\;\varid{record}\;\conid{Input}\;\keyword{where}\;\ldots{}\<[E]%
\\[\blanklineskip]%
\>[B]{}\keyword{instance}\;({}\<[13]%
\>[13]{}\ldots{}\<[165]%
\>[165]{},\conid{GetFieldFactSort}\;\varid{record}\;\varid{name}\,\sim\,\conid{Optional})\Rightarrow {}\<[E]%
\\
\>[B]{}\hsindent{5}{}\<[5]%
\>[5]{}\conid{HasFact\char95 }\;\varid{name}\;\varid{record}\;\conid{Optional}\;\keyword{where}{}\<[E]%
\\
\>[B]{}\hsindent{3}{}\<[3]%
\>[3]{}\varid{getFact}\;\varid{record}\mathrel{=}\keyword{case}\;\varid{getRecordField}\;@\varid{name}\;\varid{record}\;\keyword{of}{}\<[E]%
\\
\>[3]{}\hsindent{2}{}\<[5]%
\>[5]{}\conid{AlreadyKnown}\;\varid{result}\to \varid{return}\;\varid{result}{}\<[E]%
\\
\>[3]{}\hsindent{2}{}\<[5]%
\>[5]{}\conid{MustDerive}\to \keyword{case}\;\varid{rule}\;@\varid{record}\;@\varid{name}\;\varid{key}\;\keyword{of}{}\<[E]%
\\
\>[5]{}\hsindent{2}{}\<[7]%
\>[7]{}\conid{NoRule}\to \varid{missingRule}\;@\varid{name}\;\varid{record}{}\<[E]%
\\
\>[5]{}\hsindent{2}{}\<[7]%
\>[7]{}\conid{HasRule}\;\varid{r}\to \keyword{do}{}\<[E]%
\\
\>[7]{}\hsindent{2}{}\<[9]%
\>[9]{}\varid{result}\leftarrow \varid{runComputationRule}\;\varid{r}{}\<[E]%
\\
\>[7]{}\hsindent{2}{}\<[9]%
\>[9]{}\varid{updateDatabase}\;\varid{key}\mathbin{\$}\lambda \varid{rec}\to {}\<[E]%
\\
\>[9]{}\hsindent{2}{}\<[11]%
\>[11]{}\varid{setRecordField}\;@\varid{name}\;\varid{rec}\;(\conid{AlreadyKnown}\;\varid{result}){}\<[E]%
\\
\>[7]{}\hsindent{2}{}\<[9]%
\>[9]{}\varid{return}\;\varid{result}{}\<[E]%
\\
\>[3]{}\hsindent{2}{}\<[5]%
\>[5]{}\keyword{where}{}\<[E]%
\\
\>[5]{}\hsindent{2}{}\<[7]%
\>[7]{}\varid{key}\mathrel{=}\varid{getRecordField}\;@\text{\ttfamily \char34 key\char34}\;\varid{record}{}\<[E]%
\ColumnHook
\end{hscode}\resethooks

The \ensuremath{\conid{HasFact\char95 }} class is actually quite simple: it allows retrieval
of a fact from a record in the \ensuremath{\conid{EigerM}} monad.
The first two instances declared are also pretty easy. We have a
special case for the \ensuremath{\text{\ttfamily \char34 key\char34}} field, which is declared outside the
\ensuremath{\conid{Fact}} structure, as it is required for every Eiger record type.
\ensuremath{\conid{Input}} fields just return their contents.

For \ensuremath{\conid{Optional}} fields, we first check whether the field value is
\ensuremath{\conid{AlreadyKnown}}; if so, return it. Otherwise, we look up the rule.
The constraints to this instance insist that there is an \ensuremath{\keyword{instance}\;\conid{Rule}\;\varid{record}\;\varid{name}}
available, but it is possible that this instance specifies \ensuremath{\conid{NoRule}}.
If there is \ensuremath{\conid{NoRule}},
we throw an error in the \ensuremath{\conid{EigerM}} monad using \ensuremath{\varid{missingRule}}.
Otherwise, run the rule, updating the database with the result
(to avoid recomputation).


\subsubsection{JSON serialization via run-time instance lookup}

We must provide users a way of inputting a large quantity of data to seed the database, and of outputting an updated database after performing computations.
\Eiger/ thus supports serialization and deserialization with JSON.
The design for such a mechanism is challenging, especially for deserialization: the library does not know the set of types that need to be parsed, and the input will be heterogeneous.
Because an input may have an arbitrary number of different records of different types, each one must declare its own type.
Here is an example of what an input JSON file might look like:
\begin{hscode}\SaveRestoreHook
\column{B}{@{}>{\hspre}c<{\hspost}@{}}%
\column{BE}{@{}l@{}}%
\column{4}{@{}>{\hspre}l<{\hspost}@{}}%
\column{13}{@{}>{\hspre}c<{\hspost}@{}}%
\column{13E}{@{}l@{}}%
\column{16}{@{}>{\hspre}l<{\hspost}@{}}%
\column{21}{@{}>{\hspre}c<{\hspost}@{}}%
\column{21E}{@{}l@{}}%
\column{24}{@{}>{\hspre}l<{\hspost}@{}}%
\column{E}{@{}>{\hspre}l<{\hspost}@{}}%
\>[B]{}\{\mskip1.5mu {}\<[BE]%
\>[4]{}\text{\ttfamily \char34 Corp\char34}\mathbin{:}{}\<[13]%
\>[13]{}\{\mskip1.5mu {}\<[13E]%
\>[16]{}\text{\ttfamily \char34 type\char34}\mathbin{:}\text{\ttfamily \char34 Entity\char34}{}\<[E]%
\\
\>[13]{},{}\<[13E]%
\>[16]{}\text{\ttfamily \char34 jurisdiction\char34}\mathbin{:}\text{\ttfamily \char34 Switzerland\char34}{}\<[E]%
\\
\>[13]{},{}\<[13E]%
\>[16]{}\text{\ttfamily \char34 fiscal\char95 year\char34}\mathbin{:}\mathrm{2022}{}\<[E]%
\\
\>[13]{},{}\<[13E]%
\>[16]{}\text{\ttfamily \char34 stock\char95 based\char95 compensation\char34}\mathbin{:}\mathrm{12345.00}\mskip1.5mu\}{}\<[E]%
\\
\>[B]{},{}\<[BE]%
\>[4]{}\text{\ttfamily \char34 Switzerland\char34}\mathbin{:}{}\<[21]%
\>[21]{}\{\mskip1.5mu {}\<[21E]%
\>[24]{}\text{\ttfamily \char34 type\char34}\mathbin{:}\text{\ttfamily \char34 Jurisdiction\char34}{}\<[E]%
\\
\>[21]{},{}\<[21E]%
\>[24]{}\text{\ttfamily \char34 fiscal\char95 year\char34}\mathbin{:}\mathrm{2022}{}\<[E]%
\\
\>[21]{},{}\<[21E]%
\>[24]{}\text{\ttfamily \char34 top\char95 up\char95 tax\char95 percentage\char34}\mathbin{:}\mathrm{0.03}\mskip1.5mu\}\mskip1.5mu\}{}\<[E]%
\ColumnHook
\end{hscode}\resethooks
The top-level structure is a JSON object, where the keys of the object will turn into keys in the \ensuremath{\conid{Database}}.
Each object in the input file must include its \ensuremath{\text{\ttfamily \char34 type\char34}}: this field then tells us how to interpret the remaining fields, which are all assumed to be \ensuremath{\conid{Fact}}s in the relevant Haskell type declaration.

While that sounds simple enough, recall that this is Haskell, a statically typed language.
We cannot normally read in the name of a type as a string and then proceed to parse and produce an object of that type.
To do so, we need a run-time description of the type (including its name) and an ability to use this description to produce an actual value of that type.
Typically, this ``type description'' can be encapsulated in an instance dictionary, perhaps of the \ensuremath{\conid{FromJSON}} class.
Yet recall that we get the actual choice of type at \emph{run-time}, so typical patterns of instance constraints will not suffice.

Instead, in order to parse and produce an \eiger/ \ensuremath{\conid{Database}}, we must be able to look up instances at run-time.
Despite sounding exotic, it turns out that this is relatively straightforward. Here is the definition:%
\footnote{In the real implementation, we have a further challenge: should type names be qualified or not?
And if they are qualified, should the qualification include the package or just the modules?
The implementation handles these concerns by generalizing the interface presented here to be polymorphic over the choice of type naming discipline.
For simplicity in this text, we assume types are printed unqualified.}
\begin{hscode}\SaveRestoreHook
\column{B}{@{}>{\hspre}l<{\hspost}@{}}%
\column{3}{@{}>{\hspre}l<{\hspost}@{}}%
\column{E}{@{}>{\hspre}l<{\hspost}@{}}%
\>[B]{}\keyword{type}\;\conid{Instances}\mathbin{::}\keyword{$\forall$} \hsforall \;\varid{k}\hsdot{.}{.\,}(\varid{k}\to \conid{Constraint})\to \conid{Type}{}\<[E]%
\\
\>[B]{}\keyword{newtype}\;\conid{Instances}\;\varid{c}\mathrel{=}\conid{MkInstances}\;(\conid{Map}\;\conid{String}\;(\conid{EDict}\;\varid{c})){}\<[E]%
\\[\blanklineskip]%
\>[B]{}\keyword{type}\;\conid{EDict}\mathbin{::}\keyword{$\forall$} \hsforall \;\varid{k}\hsdot{.}{.\,}(\varid{k}\to \conid{Constraint})\to \conid{Type}{}\<[E]%
\\
\>[B]{}\keyword{data}\;\conid{EDict}\;\varid{c}\;\keyword{where}{}\<[E]%
\\
\>[B]{}\hsindent{3}{}\<[3]%
\>[3]{}\conid{PackDict}\mathbin{::}\keyword{$\forall$} \hsforall \;\varid{x}\;\varid{c}\hsdot{.}{.\,}\varid{c}\;\varid{x}\Rightarrow \conid{EDict}\;\varid{c}{}\<[E]%
\ColumnHook
\end{hscode}\resethooks

An \ensuremath{\conid{Instances}\;\varid{c}} maps from the name of a type \ensuremath{\tau} to an instance \ensuremath{\varid{c}\;\tau}, as packed in an \ensuremath{\conid{EDict}}.
The \ensuremath{\conid{EDict}} type captures an instance as a constraint on a data constructor, hiding the concrete type \ensuremath{\varid{x}} of the argument to the class \ensuremath{\varid{c}}.
The type system here does not ensure that the key in a mapping is the string representation of the type packed in the value \ensuremath{\conid{EDict}}; we will have to check for this when we retrieve instances from this database.

We can create a singleton \ensuremath{\conid{Instances}} with \ensuremath{\varid{instanceForTypeRep}}:
\begin{hscode}\SaveRestoreHook
\column{B}{@{}>{\hspre}l<{\hspost}@{}}%
\column{3}{@{}>{\hspre}l<{\hspost}@{}}%
\column{E}{@{}>{\hspre}l<{\hspost}@{}}%
\>[B]{}\varid{instanceForTypeRep}\mathbin{::}\keyword{$\forall$} \hsforall \;\varid{x}\;\varid{c}\hsdot{.}{.\,}\varid{c}\;\varid{x}\Rightarrow \conid{TypeRep}\;\varid{x}\to \conid{Instances}\;\varid{c}{}\<[E]%
\\
\>[B]{}\varid{instanceForTypeRep}\;\varid{tr}\mathrel{=}{}\<[E]%
\\
\>[B]{}\hsindent{3}{}\<[3]%
\>[3]{}\conid{MkInstances}\;(\varid{singleton}\;(\varid{show}\;\varid{tr})\;(\conid{PackDict}\;@\varid{x})){}\<[E]%
\ColumnHook
\end{hscode}\resethooks
From there, because \ensuremath{\conid{Instances}} is a \ensuremath{\conid{Monoid}}, we can make an arbitrary collection of instances.

Retrieval is a little more awkward, because Haskell does not permit a function to return a class constraint:
\begin{hscode}\SaveRestoreHook
\column{B}{@{}>{\hspre}l<{\hspost}@{}}%
\column{3}{@{}>{\hspre}l<{\hspost}@{}}%
\column{5}{@{}>{\hspre}l<{\hspost}@{}}%
\column{E}{@{}>{\hspre}l<{\hspost}@{}}%
\>[B]{}\varid{withInstance}\mathbin{::}\conid{Instances}\;\varid{c}\to \conid{String}\to {}\<[E]%
\\
\>[B]{}\hsindent{3}{}\<[3]%
\>[3]{}(\keyword{$\forall$} \hsforall \;\varid{x}\hsdot{.}{.\,}\varid{c}\;\varid{x}\Rightarrow \conid{Proxy}\;\varid{x}\to \varid{r})\to \conid{Maybe}\;\varid{r}{}\<[E]%
\\
\>[B]{}\varid{withInstance}\;(\conid{MkInstances}\;\varid{mapping})\;\varid{type\char95 name}\;\varid{f}\mathrel{=}{}\<[E]%
\\
\>[B]{}\hsindent{3}{}\<[3]%
\>[3]{}\varid{lookup}\;\varid{type\char95 name}\;\varid{mapping}\mathop{{\langle}{\&}{\rangle}}{}\<[E]%
\\
\>[3]{}\hsindent{2}{}\<[5]%
\>[5]{}\lambda (\conid{PackDict}\;\!\;@\varid{x})\to \varid{f}\;(\conid{Proxy}\;@\varid{x}){}\<[E]%
\\
\>[B]{}\mbox{\onelinecomment  \ensuremath{\mathop{{\langle}{\&}{\rangle}}} is \ensuremath{\varid{flip}\;\varid{fmap}}, used because \ensuremath{\varid{lookup}} returns a \ensuremath{\conid{Maybe}}}{}\<[E]%
\ColumnHook
\end{hscode}\resethooks
The \ensuremath{\varid{withInstance}} function must take a continuation that describes what to do with the retrieved instance.
Furthermore, it must use a \ensuremath{\conid{Proxy}} to pass the retrieved type \ensuremath{\varid{x}}, as GHC does not yet support binding type variables in higher-rank contexts~\cite{type-variables-proposal}.
Note that the function \ensuremath{\varid{f}} requires a \ensuremath{\varid{c}\;\varid{x}} constraint; this constraint is available because of the pattern-match on \ensuremath{\conid{PackDict}}, which stored the constraint evidence.

To make \ensuremath{\conid{Instances}} usable in practice, though, we cannot expect users to write out many invocations of \varid{in\-stance\-For\-Type\-Rep} and combine the results.
Instead, we provide the following Template Haskell function:
\begin{hscode}\SaveRestoreHook
\column{B}{@{}>{\hspre}l<{\hspost}@{}}%
\column{E}{@{}>{\hspre}l<{\hspost}@{}}%
\>[B]{}\varid{allGroundInstances}\mathbin{::}\conid{Q}\;\conid{Type}\to \conid{Q}\;\conid{Exp}{}\<[E]%
\\[\blanklineskip]%
\>[B]{}\mbox{\onelinecomment  example usage:}{}\<[E]%
\\
\>[B]{}{\$}\varid{allGroundInstances}\;[\varid{t}|\;\conid{FromJSON}\;|]{}\<[E]%
\ColumnHook
\end{hscode}\resethooks
The \ensuremath{\varid{allGroundInstances}} function looks for all instances of the requested class -- \ensuremath{\conid{FromJSON}} in the example -- and creates an \ensuremath{\conid{Instances}} that stores all of them.
By ``ground'' here, we mean that the instance must not contain any variables.
For \eiger/ records, which are not polymorphic, this is a reasonable restriction.
Neither \ensuremath{\varid{allGroundInstances}} nor any of the \ensuremath{\conid{Instances}} machinery works with polymorphic instances, though it would not be hard to imagine extending them to do so, with the help of an algorithm for matching template types against concrete ones. 
This extra feature was not necessary in the \eiger/ context and so remains as future work.

Powered by \ensuremath{\conid{Instances}}, the serializer and deserializer have the usual complexities of real-world software.
Both are implemented through the use of generic programming and the \package{generics-sop} library~\cite{generics-sop-paper}, looking up instances via several \ensuremath{\conid{Instances}} databases as they go.

\subsection{The interface for authors}

All of the frantic paddling under the surface supports a simple interface for authors (in the sense of \pref{sec:workflow}), who we envision to be intermediate Haskell programmers working alongside a domain expert in the regulation being encoded.
The last stage of our tour through \eiger/ highlights the design choices that make for an easier user interface.

\subsubsection{Template Haskell}

In order for our high-power implementation to work, the \eiger/ record types must implement several different classes.
This set starts with \ensuremath{\conid{Generic}} (from \ensuremath{\conid{\conid{GHC}.Generics}}) but also includes several from \package{generic-sop} and beyond.
In order to simplify the process of declaring an \eiger/ record, programmers must write only \ensuremath{\keyword{deriving}} \ensuremath{\conid{IsRecord}} after their declarations (using the \ext{DeriveAnyClass} extension).
(The \ensuremath{\conid{IsRecord}} class has no methods; it serves only as a marker to be used in this mechanism.)
Then, separately, users write \ensuremath{{\$}\varid{deriveRecordInstances}}, which looks for all \ensuremath{\conid{IsRecord}} instances to find the set of \eiger/ record types and then derives all the other instances necessary.
Using Template Haskell to write these instances also creates a nice abstraction, because it means that if \eiger/ needs new instances in the future, all we need to update is \varid{de\-rive\-Re\-cord\-In\-stan\-ces}; users can remain blissfully unaware of this.

\subsubsection{\ensuremath{\conid{Rule}}}

The \ensuremath{\conid{Rule}} class instantiated in \pref{sec:writing-rules} is defined like this:
\begin{hscode}\SaveRestoreHook
\column{B}{@{}>{\hspre}l<{\hspost}@{}}%
\column{3}{@{}>{\hspre}l<{\hspost}@{}}%
\column{5}{@{}>{\hspre}l<{\hspost}@{}}%
\column{22}{@{}>{\hspre}c<{\hspost}@{}}%
\column{22E}{@{}l@{}}%
\column{25}{@{}>{\hspre}l<{\hspost}@{}}%
\column{E}{@{}>{\hspre}l<{\hspost}@{}}%
\>[B]{}\keyword{class}\;\conid{Rule}\;\varid{record}\;\varid{field\char95 name}\;\keyword{where}{}\<[E]%
\\
\>[B]{}\hsindent{3}{}\<[3]%
\>[3]{}\varid{rule}\mathbin{::}\conid{HasRules}\;\varid{record}\Rightarrow \conid{Key}\;\varid{record}\to {}\<[E]%
\\
\>[3]{}\hsindent{2}{}\<[5]%
\>[5]{}\conid{OptionalRule}\;(\conid{FactType}\;\varid{record}\;\varid{field\char95 name}){}\<[E]%
\\[\blanklineskip]%
\>[B]{}\keyword{type}\;\keyword{family}\;\conid{HasRules}\;\varid{record}\mathbin{::}\conid{Constraint}{}\<[E]%
\\[\blanklineskip]%
\>[B]{}\keyword{data}\;\conid{OptionalRule}\;\varid{t}{}\<[22]%
\>[22]{}\mathrel{=}{}\<[22E]%
\>[25]{}\conid{NoRule}{}\<[E]%
\\
\>[22]{}\mid {}\<[22E]%
\>[25]{}\conid{HasRule}\;(\conid{ComputationRule}\;\varid{t}){}\<[E]%
\ColumnHook
\end{hscode}\resethooks
As we have seen, authors write \ensuremath{\conid{Rule}} instances to define the rules for \ensuremath{\conid{Optional}} facts.
Note that \ensuremath{\varid{rule}} returns an \ensuremath{\conid{OptionalRule}}, meaning that users might decide not to provide a rule for a given fact; this would be used when a dataset might omit a fact, but there is still no rule to compute the missing fact from other facts.

The \ensuremath{\varid{rule}} function also has a constraint \ensuremath{\conid{HasRules}\;\varid{record}}. This constraint, defined via a type family, includes \ensuremath{\conid{Rule}} constraints for all \ensuremath{\conid{Optional}} facts in an \eiger/ record type.
For example, if we have
\begin{hscode}\SaveRestoreHook
\column{B}{@{}>{\hspre}l<{\hspost}@{}}%
\column{3}{@{}>{\hspre}l<{\hspost}@{}}%
\column{E}{@{}>{\hspre}l<{\hspost}@{}}%
\>[B]{}\keyword{data}\;\conid{Example}\mathrel{=}\conid{MkEx}\;\{\mskip1.5mu \varid{ex1},\varid{ex2}\mathbin{::}\conid{Fact}\;\conid{Optional}\;\conid{Integer}\mskip1.5mu\}{}\<[E]%
\\
\>[B]{}\hsindent{3}{}\<[3]%
\>[3]{}\keyword{deriving}\;\conid{IsRecord}{}\<[E]%
\ColumnHook
\end{hscode}\resethooks
then \ensuremath{\conid{HasRules}\;\conid{Example}} would expand to \ensuremath{(\conid{Rule}\;\conid{Example}\;\text{\ttfamily \char34 ex1\char34},} \ensuremath{\conid{Rule}\;\conid{Example}\;\text{\ttfamily \char34 ex2\char34})}.
The purpose of \ensuremath{\conid{HasRules}} is to make sure that the body of the rule is able to refer to all other fields in the same record.
Without \ensuremath{\conid{HasRules}}, authors would have to ensure that all \ensuremath{\conid{Rule}} instances for an \eiger/ record type are in the same module; with \ensuremath{\conid{HasRules}}, they can be distributed across modules.%
\footnote{This treatment does not extend to fields of other record types, whose \ensuremath{\conid{Rule}} instances must be in scope in order to access the fields.
It is possible to expand the \ensuremath{\conid{HasRules}} mechanism to cover this pattern if necessary.}

Another possible design is to denote whether or not a fact has an associated rule in its type, instead of through the \ensuremath{\conid{NoRule}}/\ensuremath{\conid{HasRule}} constructors.
There does not seem to be a concrete advantage to doing so, but as we gain experience with \eiger/, it is possible this design will evolve.

\subsubsection{\ext{OverloadedRecordDot}}
\label{sec:get-field}
\label{sec:overloaded-record-dot}

We wish authors to be able to use familiar dot-projection syntax, like \ensuremath{\varid{self}\hsdot{.}{.\,}\varid{tax\char95 rate}}.
GHC's recent \ext{OverloadedRecordDot}~\cite{overloaded-record-dot-proposal} extension permits this: an expression \ensuremath{\varid{r}\hsdot{.}{.\,}\varid{f}} expands to a call to \ensuremath{\varid{getField}\;@\text{\ttfamily \char34 f\char34}\;\varid{r}}.
However, the built-in interpretation of \ensuremath{\varid{getField}} (\S\ref{sec:has-field}) is not powerful enough to work with
\ensuremath{\conid{ComputationRule}}s.
Accordingly, we use \ext{OverloadedRecordDot} with \ext{RebindableSyntax}.

We must now provide our own \ensuremath{\varid{getField}} function.
As \ensuremath{\varid{getField}} must work at many different types, it must be a class method.
The definition of this class follows:
\begin{hscode}\SaveRestoreHook
\column{B}{@{}>{\hspre}l<{\hspost}@{}}%
\column{3}{@{}>{\hspre}l<{\hspost}@{}}%
\column{E}{@{}>{\hspre}l<{\hspost}@{}}%
\>[B]{}\keyword{type}\;\conid{GetField}\mathbin{::}\conid{Symbol}\to \conid{Type}\to \conid{Constraint}{}\<[E]%
\\
\>[B]{}\keyword{class}\;\conid{GetField}\;\varid{field\char95 name}\;\varid{record}\;\keyword{where}{}\<[E]%
\\
\>[B]{}\hsindent{3}{}\<[3]%
\>[3]{}\keyword{type}\;\conid{FieldAccessResult}\;\varid{field\char95 name}\;\varid{record}\mathbin{::}\conid{Type}{}\<[E]%
\\
\>[B]{}\hsindent{3}{}\<[3]%
\>[3]{}\varid{getField}\mathbin{::}\varid{record}\to \conid{FieldAccessResult}\;\varid{field\char95 name}\;\varid{record}{}\<[E]%
\ColumnHook
\end{hscode}\resethooks
\Eiger/ includes instances to look names up in both \ensuremath{\conid{Key}\;\varid{record}} and \ensuremath{\conid{ComputationRule}\;(\conid{Key}\;\varid{record})} to flexibly allow users to project both from raw keys (such as the \ensuremath{\varid{self}} in a \ensuremath{\conid{Rule}} definition) and rules to produce keys (such as an expression like \ensuremath{\varid{self}\hsdot{.}{.\,}\varid{jurisdiction}}, when \ensuremath{\varid{self}} is a \ensuremath{\conid{Key}\;\conid{Entity}}).
Writing these (elided) instances poses no challenge.

%





\subsubsection{DSL}

To allow authors to work at the level of abstraction they understand best, the \eiger/ library includes a module \ensuremath{\conid{\conid{Eiger}.DSL}} that exports functions that work exclusively over \ensuremath{\conid{ComputationRule}}s.
For example, it exports the following:
\begin{hscode}\SaveRestoreHook
\column{B}{@{}>{\hspre}l<{\hspost}@{}}%
\column{3}{@{}>{\hspre}l<{\hspost}@{}}%
\column{E}{@{}>{\hspre}l<{\hspost}@{}}%
\>[B]{}(\mathop{==})\mathbin{::}\conid{Eq}\;\varid{a}\Rightarrow \conid{ComputationRule}\;\varid{a}\to \conid{ComputationRule}\;\varid{a}\to {}\<[E]%
\\
\>[B]{}\hsindent{3}{}\<[3]%
\>[3]{}\conid{ComputationRule}\;\conid{Bool}{}\<[E]%
\\[\blanklineskip]%
\>[B]{}\mbox{\onelinecomment  Used by the \ensuremath{\keyword{if}} \ldots \ensuremath{\keyword{then}} \ldots \ensuremath{\keyword{else}} construct}{}\<[E]%
\\
\>[B]{}\varid{ifThenElse}\mathbin{::}\conid{ComputationRule}\;\conid{Bool}\to \conid{ComputationRule}\;\varid{a}\to {}\<[E]%
\\
\>[B]{}\hsindent{3}{}\<[3]%
\>[3]{}\conid{ComputationRule}\;\varid{a}\to \conid{ComputationRule}\;\varid{a}{}\<[E]%
\\[\blanklineskip]%
\>[B]{}\mbox{\onelinecomment  inject ordinary values into a \ensuremath{\conid{ComputationRule}}}{}\<[E]%
\\
\>[B]{}\varid{use}\mathbin{::}\varid{a}\to \conid{ComputationRule}\;\varid{a}{}\<[E]%
\\
\>[B]{}\varid{use}\mathrel{=}\varid{pure}{}\<[E]%
\ColumnHook
\end{hscode}\resethooks
In addition to these, numerical instances are provided for \ensuremath{\conid{ComputationRule}}, so two \ensuremath{\conid{ComputationRule}}s can be combined by \ensuremath{\mathbin{+}}, for example.
This makes for a smooth user interface, where \emph{everything} is a \ensuremath{\conid{ComputationRule}}.

\paragraph{Conclusion}
We have now toured the highlights of the \eiger/ implementation.
One of its distinctive qualities is how it works hard under the surface, using lots of fancy types and footwork, but provides a simple interface to users, putting the user experience above all.

\section{Related Work}
\label{sec:related}
There has been a trend over the past decade of bringing digitalization and automation to the domain of legislation and regulation.
The French Public Finances Directorate has been pioneering this space by encoding the French Tax Code in a DSL and publishing parts of it for the purpose of explainability \cite{french-tax-code}.
The OECD has been running a Rules-as-Code program that ``proposes to create a machine-consumable version of some types of government rules, to exist alongside the existing natural language counterpart'' \cite{oecd-cracking-the-code}.
And the European Commission has recently had a project on machine-readable and executable reporting requirements in the financial services industry \cite{mrer-tender}.

Several new pieces of technology have been introduced in this field.
Catala is a new, text-based  DSL \cite{catala}, while Blawx is a new, graphical DSL \cite{blawx}.
These efforts are bringing natural languages and programming languages closer, coming from the direction of the latter.
In contrast, so-called controlled natural languages are closing this gap coming from the other direction by restricting the grammar and vocabulary of natural languages in order to eliminate ambiguity.
RegelSpraak \cite{regelspraak} is an implementation of this idea pursued by the Dutch Tax and Customs Administration (DTCA).
Blurring the line between natural and programming language is the Agile Law Execution Factory, also developed by the DTCA \\ \cite{alef}.
It is based on JetBrain's Meta Programming System\footnote{\url{https://www.jetbrains.com/mps/}}, a framework and IDE for building new DSLs.

These new technologies are exciting to see, and they are solving relevant problems.
They also require considerable investment in the development of the ecosystem necessary for mainstream, commercial use.
We are thus pursuing the approach of an embedded DSL, giving us a head-start towards commercial adoption.
We are also explicit about the methodology that we believe will work in practice, which is pairing engineers with domain experts, as opposed to expecting domain experts to undergo training to write high quality code.

Regarding the run-time of \eiger/, there are similarities with build systems~\cite[e.g.,][]{build-systems-a-la-carte}, self-adjusting computing~\cite{self-adjusting-computation}, as well as graph processing \cite{pregel,chromatic-scheduling}.
As we scale, we will have to address performance bottlenecks as they appear.
But for now it seems that the computations performed by \eiger/ rules are so trivial that we do not expect to be able to amortize the overhead of tracking dependencies and changes.

\section{Future Work}
\label{sec:future}

We are currently planning to use \eiger/ to engage clients, putting the ideas in this paper into practice,
along the lines of our outline in \pref{sec:workflow}.
We expect to use \eiger/ both internally for our own analysis and embedded in SaaS products for our clients.
We expect \eiger/ to be applicable and beneficial to all the tax and legal services we offer today.
Indeed, the enthusiasm for \eiger/ within \ifanon our professional services consultancy \else PwC Switzerland\fi is such that we wonder whether \eiger/ -- or a tool inspired by it -- will become a ``killer app'' for Haskell.

Longer term, we envision working with authorities to have them draft and publish authoritative \eiger/ implementations of new regulations.
We believe this is where the biggest opportunities of this technology lie for society as a whole.

On a technical level, there is still much we would like to build.
For example, in the VAT domain, the typical client has several millions transactions per reporting period that need to be processed, so performance work will be a focus.
The numerical hierarchy we have introduced allows us to, say, multiply a \ensuremath{\conid{Percentage}} and a \ensuremath{\conid{EuroAmount}} (while preventing us from multiplying two \ensuremath{\conid{EuroAmount}}s), but does it scale beyond that?
We will learn by experience.
Along similar lines, we have included a few easy consultant-facing interactions, but we recognize more will be necessary for a full-fledged product.
Nevertheless, our experience with \eiger/ as it exists -- and the ease with which we have already extended it from its core -- suggests that growth will be smooth.

In the end, we expect usage of \eiger/ to have a marked impact on \ifanon our professional services consultancy\else PwC Switzerland\fi.
Our commitment to pair-pro\-gram\-ming has already led to successes, and the type discipline imposed by working in Haskell has helped us to understand regulation more deeply.
We look forward to deploying \eiger/ and using it to deliver value to our clients.









\bibliographystyle{ACM-Reference-Format}
\bibliography{bib}

\ifextended
\appendix

\section{Feedback of the User Study}
Here we share the verbatim responses of the user study presented in this paper.
Please refer to \pref{sec:user-study} for the related questions.
\subsection{BEPS}
\begin{enumerate}
\item ``After the second session, I felt comfortable with the approach as we were also able to cover more rules.
I think it is an experience that one can get used to after a short amount of time.
Also as the code focuses mainly on the mathematical approach, an in-depth knowledge of the underlying material was not necessary.
Also having a small group helped to feel comfortable with the approach.''
\item ``It generally takes one or two rules to get into the flow during a session.
Once in the flow, the sessions are really productive.
Thus, I found it helpful when the session were a little longer.''
\item ``I am confident that we implemented our faithful interpretation of the legal text.
As it is a complex matter that is still subject to a lot of interpretation and clarification, as well as outstanding guidance, the code most probably will become subject to change.
We have also focused on the less complex rules thus far.''
\item ``Now that I'm familiar with the structure, I feel that I would be able to read someone else's implementation after been given a short introduction.''
\item ``Currently, I would still use the legal text as it is nicer to present.
Depending on how the final tool will look, I could consider showing the code, especially when it comes to the calculation part which is shown in a more intuitive way.''
\item ``For me, as a personal investment, I found it really helpful to understand the rules from a more holistic and mathematical angle.
Connections between rules are also becoming more obvious.
I can also see the benefits of a code that can be provided to clients to perform their own calculations.
[...]''
\item ``I think the code is really helpful when it comes to the mathematical application.
Informative text in the regulation is not reflected in the code, which still needs to be communicated to the client in another way.''
\item ``So far the process has been really informative for my benefit. Moreover, it is easy to work with [the engineer].
I always had time to think through a rule or change it and I never felt rushed or incompetent.
In my point of view, to structure and organize the project/process more, I would suggest to have a timeline/going to market strategy including milestones prepared in order for everyone to have a clear goal/deadline insight.''
\end{enumerate}

\subsection{CBAM}
\begin{enumerate}
\item ``The step-by-step, article-by-article approach makes it easy to follow, to understand, and ultimately to implement the coding.''
\item ``Seeing the result of the code straight away would be interesting.
If, for instance, some dummy data was used, what would the result of the code be? Would it result in an error, in an incorrect result, or the correct result?''
\item ``In my view, the CBAM text is relatively simple -- articles are relatively short and there are not many cross-references to articles.
I have high confidence that the legal text is faithfully implemented.''
\item ``Reading and understanding other people's code can be a challenge.
Assuming, however, that other people follow a logical and structured way of coding, I have high confidence that I could read and understand their implementation of legal text.''
\item ``My preference would be to use the code but also include the legal text as non-executable comments to provide context.''
\item ``The pair-programming approach is essential, particularly in the beginning.
Coders need to get an understanding of the legal text's logic, and subject matter experts (SMEs) need to understand the logic of the code.
The benefits goes both ways.
I actually do believe that SMEs benefit most.
Dissecting the legal text in code-able parts may also help SMEs to make implicit assumptions explicit, and to look at the legal text from a different perspective.''
\item ``No, I think the code could replicate the legal text without struggle.''
\item ``My overall feedback is very positive.
The process of transforming legal text to code is clear and easy to follow.
One challenge that SMEs may struggle with is the direction of logic.
Legal text often starts with a general claim, followed by specific conditions or exceptions.
Code often works the other way around.
Specifics are first defined, and more general claims are build upon this.
I therefore think it is important to include non-executable comments in the code to provide context.''
\end{enumerate}

\subsection{UK Value Added Tax Regime}
\begin{enumerate}
\item ``I found it much easier to get to grips with the approach than I had expected.
I had expected it to take some time for us to build the ground rules for a specific domain so that we could create something that worked, but I was surprised that we were already working on functional code within 5 minutes of starting.
This is one of the key benefits of pairing two people with specific domain expertise (one “software technical” and one “domain technical”) so each can focus on their specialism and speed is therefore increased.
I was surprised at how quickly I could pick up the logic of the code that was being written with my input.''
\item ``If it is possible to identify some terms we can either infer or have as default values so that the creation of rules or facts can be streamlined to a degree, that would help make the experience better.''
\item ``I am confident that what we built would faithfully represent the basic functionality of the VAT system.
There are a few edge cases that would specifically need to be incorporated for it to be fully representative of the framework, but I am confident that with a bit more coding time we would be able to represent the full set of scenarios.''
\item ``One of the benefits of the pair-programming approach is that you experience the code being built “first hand”, meaning that you understand why each term is being used.
Reviewing it “second hand” means you do not have some of this context, but having said that, the code was surprisingly easy to read even for examples where I had not been present at writing.
I would say certainly no more difficult than reading formulas in Excel, just different, and not beyond the capability of a domain expert to learn within a reasonably short timeframe.''
\item ``If not the code, what is missing for you to use this instead? When advising on regulation, we must default to the text that is legally enforceable.
At this point, this is the legal text -- particularly as the industry is built on human language.
However, with faithfully represented code like the examples here, I could see the industry moving to a place where the code could be used to advise on implementation.
It would be important to annotate the code with references and links to legal texts to aid the review of the implemented code and “bridge the gap” between the “old” and the “new” systems.''
\item ``Absolutely''
\item ``Not for the primary activities of regulation: aggregation of data, comparing to a simple rule book to derive conclusions, and comparatively simple calculations.
The code can deliver on these with very low execution risk in my view.
The main challenge will be the replication of areas of judgement; situations where option A or B could be a possibility, and the “right” choice is not necessarily clear, or is a tactical/strategic one.
Indeed, it is these questions that makes one organization's risk appetite different from others, and is the key reason for having a regulatory compliance team.
We need the ability with Eiger to accommodate these decision points, making the options clear and then letting a user make a decision.''
\item ``I really enjoyed the session, was pleasantly surprised with the speed of establishing executable code from a blank page.
I really think this framework has potential, and the process is effective at “upskilling” both developers on the workings of regulation, and domain experts on how to think like a developer.''
\end{enumerate}

\fi

\end{document}